\documentclass[12pt]{article}%
\usepackage{amsmath,amssymb,amsfonts}
\usepackage{color}
\usepackage{amsmath}
\usepackage{amsfonts}
\usepackage{graphicx}
\usepackage{slashed}%
\usepackage{subfig}

\newcommand{\ZZ}{\mbox{${\mathbb Z}$}}

\numberwithin{equation}{section}
\usepackage{hyperref}   % Doesn't always play well with other packages
\begin{document}

% \pagenumbering{roman}
\thispagestyle{empty}

%\begin{titlepage}
\begin{center}
{\huge \bf Goldstone Fermion Dark Matter
}

\vskip.1cm
\end{center}
\vskip0.2cm

\begin{center}
{\bf
Brando Bellazzini$^a$, Csaba Cs\'aki$^a$, Jay Hubisz$^b$, Jing Shao$^b$ and Philip Tanedo$^a$}  
\end{center}

\vskip 8pt

\begin{center}
$^{a}$ {\it 
Department of Physics, LEPP, Cornell University, Ithaca NY 14853}\\
% Laboratory of Elementary Particle Physics\\
% Cornell University, Ithaca, NY 14853, USA } \\
\vspace*{0.1cm}
$^{b}$ {\it
Department of Physics, Syracuse University, Syracuse, NY  13244}\\
\vspace*{0.3cm}
{\tt  b.bellazzini@cornell.edu, csaki@cornell.edu,jhubisz@physics.syr.edu, jishao@syr.edu, pt267@cornell.edu}
\end{center}

\vglue 0.3truecm

\begin{abstract}
\vskip 3pt
\noindent

We propose that the fermionic superpartner of a weak-scale Goldstone boson can be a natural WIMP candidate. 
The $p$-wave annihilation of this `Goldstone fermion' into pairs of Goldstone bosons automatically generates the correct relic abundance, whereas the XENON100 direct detection bounds are evaded due to suppressed couplings to the Standard Model. 
Further, it is able to avoid indirect detection constraints because the relevant $s$-wave annihilations are small.
The interactions of the Goldstone supermultiplet can induce non-standard Higgs decays and novel collider phenomenology.

% All of relevant properties of the Goldstone fermion  are independent of  explicit realizations and are presented in a low-energy effective framework fixed only  by supersymmetry and the non-linearly realized global symmetry. 

\end{abstract}

%\end{titlepage}
% \pagenumbering{arabic}
% This is a hack (see \pagenumbering{alph} above)
% Return to roman numbering

%%%%%%%%%%%%%%%%%%%%%%%%%%%%%%%%%%%%%%%%%%%%%%%%%%%%%%%%%%

\section{Introduction}

Cosmological observations now provide overwhelming evidence that about $20\%$ of the energy density of the universe is some unknown form of cold dark matter \cite{Larson:2010gs}. The most popular candidates are weakly interacting massive particles (WIMPs) which can produce the correct relic abundance after freeze out,
\begin{equation}
\label{abundance_intro}
\Omega_{\text{DM}} h^2\approx 0.1\frac{\mbox{pb}}{\langle\sigma v\rangle}\,.
\end{equation}
A natural candidate for WIMP dark matter arises in extensions of the Standard Model with low-scale supersymmetry (SUSY) and $R$-parity. In such models the lightest supersymmetric particle (LSP) is automatically stable and generically has mass on the order of the weak scale \cite{Jungman:1995df}. 

The `WIMP miracle' is the statement that a particle with a mass and annihilation cross section typical of the weak scale will automatically yield a relic abundance that is within a few orders of magnitude of the observed value. This paradigm has been challenged by recent direct detection searches for WIMPs. In particular, XENON100  recently set the most stringent upper limit on the spin-independent elastic WIMP--nucleon scattering cross section, $\sigma_{\text{SI}}=7.0\times 10^{-45}\text{ cm}^2 = 7.0 \times 10^{-9}\text{ pb}$, for a $50$ GeV WIMP at 90\% confidence \cite{Aprile:2011hi}. 
This large discrepancy between the necessary annihilation cross section and the direct detection bound is increasingly difficult to explain in the usual WIMP dark matter scenarios.

For example, within the minimal supersymmetric Standard Model (MSSM), one must typically tune parameters in order to explain this difference \cite{Farina:2011bh}. 
A standard approach is to consider parameters in which $\sigma_{\text{SI}}$ is suppressed below direct detection constraints.
At generic points in the parameter space this will also imply a suppressed annihilation cross section and thus a relic abundance that is too large. In order to overcome this problem one needs to assume special relations among a priori unrelated parameters in order to boost the annihilation rate.
For example, a pure bino LSP would require coannihilation (due to an accidental slepton degeneracy) or resonant annihilation to obtain the correct annihilation cross section \cite{Griest:1990kh}. Alternately, the observation that Higgsinos and winos have annihilation cross sections that are typically too large allows one to tune the LSP to be a specific combination of bino, Higgsino, and wino to generate the correct abundance \cite{ArkaniHamed:2006mb}. This `well-tempered neutralino' scenario, however, is now strongly disfavored by XENON100 \cite{Farina:2011bh}.

In light of this tension, it is natural to consider non-minimal SUSY models in which
\begin{itemize}\addtolength{\itemsep}{-0.5\baselineskip}
	\item the WIMP is a weak scale LSP,
	\item the direct detection cross section is suppressed while maintaining the correct relic abundance without any fine tuning, and
	\item the experimental prospects in near future include novel collider signatures.  
\end{itemize}
We therefore extend the MSSM by a new sector with an approximate global symmetry which is spontaneously broken in the supersymmetric limit. A natural WIMP candidate that satisfies the above criteria is the fermionic partner of the Goldstone boson which we refer to as the \textit{Goldstone fermion}, $\chi$. 
This particle can naturally sit at the bottom of the spectrum because it lives in the same chiral supermultiplet as the Goldstone boson $a$ and is thus protected by Goldstone's theorem and SUSY. Even when SUSY is broken, the Goldstone fermion can remain light with mass at or below $M_{\text{SUSY}}$ \cite{Tamvakis:1982mw,Chun:1995hc,Higaki:2011bz}. 
This scenario is a weak scale version of axino dark matter \cite{Rajagopal:1990yx}; for an early attempt containing similar elements see \cite{Mohapatra:1994}.
Similar realizations also appear in dark matter models where the LSP has a large ``singlino'' component \cite{Draper:2010ew}; such models can reproduce the mass spectrum of Goldstone fermion dark matter but do not have a limit where the global symmetry is broken while SUSY is exact. In particular the singlino dark matter effective interactions do not come from an effective low-energy K\"ahler potential as discussed in Section \ref{effective_Kahler}. Further, due to singlino--Higgsino mixing, such models typically require tuning to avoid direct detection bounds.
%
% Our (pseudo-)Goldstone fermion $\chi$ can naturally be at the bottom of the spectrum because it lives in the same chiral supermultiplet as the Goldstone boson $a$ and is thus protected by Goldstone's theorem and SUSY. Even when SUSY is broken the Goldstone fermion can remain light with mass at or below $M_{\text{SUSY}}$ \cite{Tamvakis:1982mw,Chun:1995hc,Higaki:2011bz}. 

The SUSY non-linear sigma model is a generic low-energy theory of the Goldstone supermultiplet based only on the symmetry breaking pattern \cite{Zumino:1979et}. It can be organized as an expansion in inverse powers of the symmetry breaking scale, $f$. In particular the leading order contribution to dark matter annihilation is controlled by a trilinear derivative coupling $\bar{\chi}\gamma^\mu\gamma^5\chi \partial_\mu a/f$.
If the global symmetry is anomalous with respect the SM gauge group, the Goldstone bosons will, in turn, decay to stable SM particles, $a\rightarrow gg\,,\,\gamma\gamma$. 
All the interactions can be perturbative and compatible with  gauge coupling unification if the mediators of the anomaly come in complete  GUT multiplets.
If the Goldstone fermion mass $m_\chi$ is around the weak scale and the symmetry breaking scale $f$ is around the TeV scale, then the resulting annihilation cross section is automatically in the thermal WIMP range 
\begin{equation}
\langle\sigma v\rangle\approx (m_{\chi}^2/f^4)(T_{f}/m_{\chi})\approx 1\mbox{ pb}.
\end{equation} 
The freeze-out temperature $T_{f}/m_{\chi}\simeq 1/20$ is insensitive to details of the model and appears because $\chi\chi\rightarrow aa$ is a $p$-wave process.

After electroweak symmetry breaking at $v_{\text{EW}}=175\mbox{ GeV}$, the CP-even scalar component of the Goldstone chiral multiplet mixes with the Higgs boson and generates an effective $h\chi\chi$ coupling which is suppressed by $m_{\chi}v_{\text{EW}}/f^2\sim 0.01$. While standard Higgsino-like dark matter in the MSSM gives a large direct detection cross section, Goldstone fermion scattering off nuclei lies just below the XENON100 bound,
 \begin{equation}
 \sigma_{\text{SI}}\approx \left(\frac{m_{\chi}v_{\text{EW}}}{f^2}\right)^2 \sigma^{\text{MSSM}}_{\text{SI}}\approx 10^{-45}\mbox{ cm}^2\,.
 \end{equation}
Note that this suppression factor in $\sigma_{\text{SI}}$ is roughly of the same order as the suppression needed in the annihilation cross section for a standard weak-scale WIMP, % i.e.\
% \begin{align}
	$\langle\sigma v\rangle_{\text{WIMP}}\sim  $ $ \pi \alpha_{\text{weak}}^2/(100\mbox{ GeV})^2$ $\sim 150\mbox{ pb}$,
% \end{align}
to obtain the correct abundance (\ref{abundance_intro}).

Finally, Goldstone fermion dark matter has novel consequences on Higgs phenomenology at the LHC. The global symmetry requires a derivative coupling between the Goldstone boson and the Higgs boson $\sim v_{\text{EW}}/f^2(\partial a)^2 h$. 
If kinematically allowed, the Higgs boson decays into four light unflavored jets, $h\rightarrow 2a\rightarrow 4j$, with a sizeable branching ratio. This decay mode is `buried' under the QCD background. Such non-standard Higgs decays have recently been investigated in SUSY models motivated by the little hierarchy problem \cite{Bellazzini:2009xt, Luty:2010vd}. 
% The discovery of a `buried Higgs' might be possible at the LHC by means of jet substructure algorithms but only with relatively high luminosity \cite{Bellazzini:2010uk}. 
For Goldstone fermion dark matter, the Higgs might only be `partially buried' with a branching ratio of 30\% to the Standard Model.
Alternately, one can hope to discover the Goldstone boson itself by looking for the $a\rightarrow 2g$ decay. Together with the direct detection of its fermionic superpartner, such a discovery would be strong evidence that the dark matter particle emerges because of the Goldstone mechanism and SUSY.

The paper is organized as follows. We introduce the effective low-energy theory of a Goldstone supermultiplet in Section \ref{Goldstones} and extend this by including SUSY and explicit global symmetry breaking in Sections \ref{susy_breaking_sect} and \ref{sec:W:from:explicit:breaking}.
Readers who are primarily interested in dark matter phenomenology can proceed directly to Sections \ref{Abundance_sect} and \ref{sec:direct_indirect_detection}, where we review (in)direct detection prospects and calculate the relic abundance. We discuss the LHC phenomenology in Section \ref{colliderPheno}. In Appendix~\ref{app:models} we present simple models that realize this scenario. Details of the annihilation cross section calculation are given in Appendix~\ref{appAnnihilation}. Remarks on a possible Sommerfeld enhancement are presented in Appendix~\ref{app:Sommerfeld}.

\section{The Goldstone Supermultiplet}
\label{Goldstones}

We consider a supersymmetric gauge theory with a global U(1) symmetry that is broken by fields $\Psi_i$ which obtain vacuum expectation values (vevs) $f_i$. 
In the limit of unbroken supersymmetry, the theory has a massless Goldstone chiral superfield,
\begin{equation}
A= \frac{1}{\sqrt{2}}(s+i a)+\sqrt{2}\theta\chi +\theta^2 F\, .
\end{equation}
% In terms of the high-energy degrees of freedom, $\psi_i=f_i e^{q_i A/f}$, $A$ is realized linearly by
% \begin{align}
% 	A =  \sum_i \frac{q_if_i}{f}\psi_i. \label{eq:A_in_terms_of_high_energy_fields}
% \end{align}
which is the low-energy degree of freedom of the high-energy fields,
\begin{equation}
\Psi_i=f_i e^{q_i A/f}\, ,
\end{equation}
where the effective symmetry breaking scale is
\begin{equation}
 f^2=\sum_i q_i^2 f_i^2\, ,
\end{equation}
and $q_i$ is the U(1) charge of $\psi_i$.
We refer to the component fields as the \textit{Goldstone boson} $a$, the \textit{sGoldstone} $s$, and the \textit{Goldstone fermion} $\chi$. 
In models where the U(1) is a Peccei-Quinn symmetry, these are typically called the axion, saxion, and axino, respectively.
The mass of the CP-odd scalar $a$ is directly protected by the Goldstone theorem while the $s$ and $\tilde{a}$ masses are, in turn, protected by supersymmetry. 

The Goldstone boson shift symmetry acts on the chiral superfield as $A\to A+i c f$. It is thus often convenient to consider a non-linear realization of the Goldstone chiral superfield, $G=e^{A/f}$, which naturally transforms under the U(1) shift symmetry, $G\to e^{ic} G$. In the absence of explicit global symmetry breaking, this shift symmetry forbids any superpotential term involving $A$. 

\subsection{Effective K\"ahler potential}
\label{effective_Kahler}

The shift symmetry restricts the dependence of the K\"ahler potential on the Goldstone superfield to take the form
\begin{equation}
K=K(A+A^\dagger, \Phi_\text{L} ).
\end{equation}
We have written $\Phi_\text{L}$ to denote light fields which are uncharged under the global symmetry. Note this general form includes the canonical term $AA^\dagger$ which is $(A+A^\dag)^2$ up to a K\"ahler transformation. 
 
We may examine the Goldstone self-interactions by expanding the canonically normalized K\"ahler metric in inverse powers of the scale $f$:
 \begin{align}
K_{(2)}=\frac{\partial^2 K}{\partial A\partial A^\dagger}=& 1+b_1\frac{ q}{f}(A+A^\dagger)+b_2\frac{q^2}{2! f^2}(A+A^\dagger)^2+\ldots\, ,
\label{eq:Kahler:metric:A}
\end{align}
where $q$ is an reference U(1) charge of the theory. The choice of $q$ is arbitrary and irrelevant since the combination $f/q$ is invariant under charge rescaling. For simplicity we set $q=1$ henceforth.
After integrating out the auxiliary fields, the general form of the Lagrangian is
\begin{align}
\mathcal{L}=&\phantom{+} K_{(2)}(s)\left(\frac{1}{2}\partial^\mu s \partial_\mu s+\frac{1}{2}\partial^\mu a \partial_\mu a+\frac{i}{2}\chi^\dagger \bar{\sigma}^\mu\partial_\mu\chi-\frac{i}{2}\partial_\mu\chi^\dagger \bar{\sigma}^\mu\chi\right)\nonumber\\
& -\frac{1}{\sqrt{2}}K_{(3)}(s)\left(\chi^\dagger  \bar{\sigma}^\mu \chi \partial_\mu a\right)+
\frac{1}{4}\left(K_{(4)}(s)-\frac{K^2_{(3)}(s)}{K_{(2)}(s)}\right)(\chi  \chi ) (\chi^\dagger \chi^\dagger)\, ,
\end{align}
where $K_{(n)}=\partial^n K/\partial A^n$. 
% Each term in this Lagrangian admits a geometrical interpretation on a curved K\"ahler manifold \cite{Zumino:1979et}.
% In particular, $K_{(3)}$ represent the affine connection whereas $K_{(4)}-K^2_{(3)}/K_{(2)}$ is the curvature tensor.
%
Passing to four-component Dirac spinors and expanding the Lagrangian in inverse powers of $1/f$ yields,
\begin{align}
\label{Lagrangian_self}
	\mathcal{L}=&\phantom{+} \left(1+b_1\frac{\sqrt{2}}{f}s+b_2\frac{1}{f^2}s^2+\cdots\right)
	\left(\frac{1}{2}\partial^\mu s \partial_\mu s+\frac{1}{2}\partial^\mu a \partial_\mu a+\frac{i}{2}\bar{\chi}\gamma^\mu\partial_\mu\chi\right)\\
\nonumber
& +\frac{1}{2\sqrt{2}}\left(b_1\frac{1}{f}+b_2\frac{\sqrt{2}}{f^2}s+\cdots\right)
\left(\bar{\chi} \gamma^\mu\gamma^5\chi\right) \partial_\mu a+
\frac{1}{16f^2}\left(b_{2}-b_{1}^2+\cdots\right)\left[(\bar{\chi}\chi)^2 - (\bar{\chi}\gamma^5\chi)^2\right]\,.
\end{align}
% \begin{align}
% \label{Lagrangian_self}
% 	\mathcal{L}=&\phantom{+} \left(1+b_1\frac{\sqrt{2}q}{f}s+b_2\frac{q^2}{f^2}s^2+\cdots\right)
% 	\left(\frac{1}{2}\partial^\mu s \partial_\mu s+\frac{1}{2}\partial^\mu a \partial_\mu a+\frac{i}{2}\bar{\chi}\gamma^\mu\partial_\mu\chi\right)\\
% \nonumber
% & +\frac{1}{2\sqrt{2}}\left(b_1\frac{q}{f}+b_2\frac{\sqrt{2}q^2}{f^2}s+\cdots\right)
% \left(\bar{\chi} \gamma^\mu\gamma^5\chi\right) \partial_\mu a+
% \frac{1}{16}\frac{q^2}{f^2}\left(b_{2}-b_{1}^2+\cdots\right)\left[(\bar{\chi}\chi)^2 - (\bar{\chi}\gamma^5\chi)^2\right]\,.
% \end{align}
The coefficients $b_{1,2,\cdots }$ completely characterize the self-interactions of the Goldstone multiplet in the symmetric limit.
The $b_1$ coefficient is particularly important for the dark matter abundance since it controls the size of the $\chi\chi a$ vertex. The tree-level contribution to $b_1$ can be determined by comparing (\ref{eq:Kahler:metric:A}) to the canonical K\"ahler potential of the high-energy fields $\psi_i$,
\begin{equation}
K=\sum_i \Psi_i^\dagger\Psi_i=\sum_i f_i^2 e^{q_i(A+A^\dagger)/f}.
\end{equation}
Note that in the absence of explicit U(1)-breaking terms, $A$ does not get a vev and $K$ is canonically normalized with respect to the Goldstone superfield. All Goldstone self-interactions are calculable from the physical K\"ahler metric,
\begin{equation}
K_A^{\phantom{A}A^\dagger}=\frac{1}{f^2}\sum_i f_i^2 q_i^2 e^{q_i(A+A^\dagger)/f}=1+
% \frac{(A+A^\dagger)}{f}\left(\sum_{i}q_i^3 f_i^2\right)/f^2+\ldots\, . 
\frac{(A+A^\dagger)}{f^3}\left(\sum_{i}q_i^3 f_i^2\right)+\ldots\, . 
\end{equation}
In particular, the tree-level contribution to the $b_{1}$ coefficient is given by
\begin{equation}
\label{eq:b1:expression}
b_{1}=\frac{1}{f^2}\sum_{i}q_i^3 f_i^2\, .
\end{equation}
Note that $b_1$ is invariant under overall charge scaling. In simple models with just two fields $\Psi_\pm$ of opposite charge, $b_1$ is bounded, $-1\leq b_1 \leq 1$. 
In general, however, there is no such restriction on $b_1$ in theories with more fields or with dynamical U(1) breaking.

\subsection{Interactions and mixing with light fields}
\label{interactions_lightFields}

Even though the MSSM fields are uncharged under the global symmetry, they may couple to the spontaneously broken sector through higher-order terms in the K\"ahler potential. We will particularly be interested in the coupling of the Goldstone multiplet with the Higgs doublets $H_{u,d}$. Explicit symmetry breaking terms can generate superpotential couplings between the MSSM and the Goldstone sector; these are discussed in Section~\ref{sec:W:from:explicit:breaking}.

The K\"ahler potential interactions between the Higgses and the Goldstone superfield can be parameterized by expanding in $1/f$,
\begin{align}
	K=&\phantom{+}\frac{1}{f}(A+A^\dagger)(c_1 H_u H_d+\ldots+{\rm h.c.} )
	 +\frac{1}{2 f^2}(A+A^\dagger)^2(c_2 H_u H_d+\ldots +{\rm h.c.}) +\mathcal{O}(1/f^3)\,.
\end{align}
Note that the first term vanishes if there is a $\ZZ_2$ discrete symmetry $A\to -A$. The presence of such symmetry depends on the choice of UV completion. A mixing between the Higgs and the sGoldstone arises, for example, from the K\"ahler metric term 
\begin{equation}
\label{generic_Higgs}
K_{H_u}^{\phantom{H}A^\dag}=\partial^2 K/(\partial H_u \partial A^\dagger)=\frac{1}{f} c_1 H_d +\ldots
\rightarrow \frac{v_{\text{EW}}}{f} c_1\cos\beta+\ldots\, .
\end{equation}
The $c_2$ terms can also give rise to mixing if the sGoldstone also gets a VEV of order $\langle s\rangle \sim f$. 
After rotating the Higgs and sGoldtone fields, the coupling between $h$ and the Goldstone multiplet appears in the effective Lagrangian as
\begin{equation}
\label{Higgs_coupling}
\mathcal{L}_{\text{eff}}=\left[\frac{1}{2}(\partial a\partial a)+\frac{i}{2}\bar{\chi}\gamma^\mu\partial_\mu\chi \right] \left(1+b_1\frac{\sqrt{2}}{f}s+c_h \frac{ v_{\text{EW}}}{f^2} h+\ldots\right) +\ldots\, ,
\end{equation}
where $c_h$ is a function of the coefficients $c_{1,2},d_{1,2}$ and the Higgs sector mixing angles. This coupling is suppressed in the large $m_s$ limit, $c_h \rightarrow (m_h/m_s)^2$.
At this order in $q^2 v_{\text{EW}}/f^2$ there are additional Higgs doublet couplings of the form   
\begin{align}
		-\frac{i}{4f^2} c_2 \bar\chi\gamma^\mu\gamma^5\chi \left(H_u\partial_\mu H_d +\partial_\mu H_u H_d- \text{h.c.}\right)
	\end{align}
 which give rise to additional interactions of the heavy Higgses with the Goldstone fermion, but do not involve the light higgs $h$. We neglect these couplings and the mixing of the heavy Higgses with the sGoldstone. 
 
Besides the scalar mixing, there is kinetic mixing between the Higgsino and Goldstone fermion of the form
\begin{eqnarray}
{\cal L}_{\text{KM}} & = & \frac{i}{2f} \left[\left(\chi^\dagger \bar{\sigma}^{\mu} \partial_{\mu} \tilde{H}_u - \partial_{\mu} \chi^\dagger \bar{\sigma}^{\mu} \tilde{H}_u \right) (c_1 H_d+\ldots) + \text{h.c.} +  (H_u \leftrightarrow H_d) \right] \nonumber \\
&\rightarrow  & i \epsilon_{u} \chi^\dagger \bar{\sigma}^{\mu} \partial_{\mu}  {\tilde H}^0_u + i \epsilon_{d} \chi^\dagger \bar{\sigma}^{\mu} \partial_{\mu}  {\tilde H}_d^0 + \text{h.c.}
\end{eqnarray}
where $\epsilon_{u,d} \sim  v_{\text{EW}}/f$. In the case where $\mu \gg m_{\chi}$, the Goldstone Fermion has a small Higgsino component roughly given by 
$\epsilon_{u,d}\, m_{\chi}/\mu \sim  \,v_{\text{EW}}\, m_{\chi}/f\mu$.

The K\"ahler terms involving the other MSSM matter fields are typically more suppressed. Assuming minimal flavor violation to control flavor-changing neutral currents, these terms take the form
%  
% K\"ahler terms involving the MSSM matter fields are necessarily \textbf{[[do we mean `typically'?]]} more suppressed. The form allowed (assuming that the Yukawa interactions $Y_{u,d,l}$ are the only source of flavor violation, in order to ensure the suppression of flavor changing neutral currents)  is
 \begin{equation}
 K= \frac{1}{f} (A+A^\dagger ) \left( \frac{Y_u}{M_u} \bar{Q} H_u u +\frac{Y_d}{M_d} \bar{Q} H_d d +\frac{Y_l}{M_L} \bar{L} H_d e + \text{h.c.} \right).
 \end{equation}
The suppression scales $M_{u,d,l}$ are not necessarily related to the global symmetry breaking scale $f$, and can be much larger depending on the UV completion of the theory.

\section{SUSY breaking}
\label{susy_breaking_sect}

We assume that soft SUSY breaking terms which simultaneously break the U(1) global symmetry are negligible. The remaining soft terms generate an explicit sGoldstone mass, but leave the Goldstone boson massless. 
The Goldstone fermion may only get a mass from the superpotential or from $D$-terms via mixing with gauginos. For simplicity we ignore the latter possibility so that the fermion mass matrix is the second derivative matrix of the superpotential,
\begin{align}
	\left(m_{\text{fermion}}\right)_{ij} = W_{ij}.
\end{align}
%%%%%
While the superpotential terms are U(1) invariant and supersymmetric, a Goldstone fermion mass can still be induced if the vacuum is shifted from its supersymmetric value due to the presence of soft breaking terms. The U(1) invariance of the superpotential implies
\begin{equation}
\label{symm_W}
\sum_j \frac{1}{f} W_{ij} q_j f_j=-\frac{1}{f}q_i W_i = -\frac{1}{f} q_i F_i \, ,
\end{equation}
so that the Goldstone fermion $\chi= \sum_i q_if_i \psi_i/f$ is indeed a zero mode of the fermion mass matrix when none of the U(1)-charged $F$-terms obtain a vev \cite{Tamvakis:1982mw}.
The mass of the Goldstone fermion then depends on whether the U(1)-charged fields pick up $F$-terms in the presence of soft breaking terms \cite{Tamvakis:1982mw,Chun:1995hc}:
\begin{equation}
\label{Tamvakis_Wyler}
m_\chi \approx q_i\langle F_i\rangle /f.
\end{equation}
 If the superpotential has an unbroken $R$-symmetry which is left unbroken by the soft terms, then $\chi$ necessarily carries $R$-charge $-1$ and a Majorana mass is prohibited\footnote{We thank Y.~Nomura for pointing out the role of $R$-symmetry.}. In particular, soft scalar masses always preserve $R$-symmetry and hence cannot generate a Goldstone fermion mass in the $R$-symmetric case.
On the other hand, $A$-terms are holomorphic and generically break $R$-symmetries. 
Thus $A$-terms are expected to always contribute to the Goldstone fermion mass, while soft scalar masses may or may not contribute. 

The effect of $A$-terms is equivalent to the mixing between the $F$-terms between U(1)-charged fields and the SUSY breaking spurion $\langle X\rangle = F\theta^2 + \cdots$. For concreteness, we consider gravity mediation with 
$F/M_{\text{Pl}} \sim m_{\text{soft}}$.
It was recently emphasized in \cite{Cheung:2011mg} that $FF_i^\dag$ type mixing terms are always expected and will contribute a mass of order $m_{3/2}$ to the Goldstone fermion. Indeed, such mixing terms arise from higher dimensional K\"ahler terms of the form
\begin{align}
K = \sum_i Z(X,X^\dag) \Phi_i^\dag\Phi_i.\label{eq:sloperator:general}
\end{align}
% For $Z(X,X^\dag) = X+X^\dag$, the quadratic term in $(A+A^\dag)$ gives a contribution to the fermion mass on the order of $m_{3/2}$; this has been referred to as the `cosmological axino problem' \cite{Cheung:2011mg}.
% %
% This, however, can be understood simply as the effect of an $A$-term which induces a non-zero $F$-term for the U(1)-charged fields.
%
%  
Using the technique of analytic continuation into superspace \cite{Giudice:1997ni}, one may absorb $Z$ into a redefinition of the chiral superfields
\begin{align}
	\Phi \to \Phi' \equiv Z^{1/2}\left(1+\frac{\partial \ln Z}{\partial X}F\theta^2\right)\Phi . \label{eq:shifted:phi}
\end{align}
This canonically normalizes $K$ and generates soft terms that include the $A$-terms 
\begin{align}
	\Delta {\cal L}_\text{soft} = \left.\frac{\partial W}{\partial \Phi}\right|_{\Phi = \phi} Z^{-1/2}\left(-\frac{\partial \ln Z}{\partial \ln X}\frac FM\right).
\end{align}
These terms completely incorporate the mixing between $F$-terms of the form $FF_i^\dag \Phi_i$. The Goldstone fermion mass is determined by the induced $F_i$s obtained by minimizing the scalar potential,
\begin{equation}
V= \left|\frac{\partial W}{\partial \phi_i}\right|^2 + A_i \frac{ \partial W}{\partial \phi_i} \phi_i+{\rm h.c.} + m_i^2 |\phi_i|^2.
\end{equation}
% Thus the fermion mass induced by Planck-suppressed terms (\ref{eq:sloperator:general})  is simply a manifestation of statement that soft breaking $A$-terms induce $F$-terms for the U(1) charged fields. The soft breaking scalar masses also induce masses of order $m_i$ to the sGoldstone boson which may also mix with other massive singlet states in the theory.

To summarize this section, we find that $A$-terms will always contribute to the Goldstone fermion mass. Assuming that $A_i, m_i < f_i$ for all $i$, the generic size of the induced $F$-terms is $|F_i| \approx A_i f_i$
% \begin{equation}
% |F_i| \approx A_i f_i\, ,
% \end{equation}
and, consequently, the induced Goldstone fermion mass is $\sim A_i q_i$. 
In many situations the $A$-terms can be suppressed relative to other soft breaking terms and it is thus reasonable to expect that the Goldstone fermion remains lighter than the other superpartners. 
Soft scalar masses may also contribute. If they do, their contribution to the $F_i$ is expected to be of order $F_i\sim m_i^2$ so that the contribution to the Goldstone fermion mass is of order $\sim m_i^2/f_i$, which again can easily be suppressed.

\section{Superpotential terms from explicit breaking}
\label{sec:W:from:explicit:breaking}
 
The shift symmetry forbids any superpotential for the Goldstone chiral multiplet $A$. 
In order to generate a small Goldstone boson mass one must include terms which break the global symmetry. These can come from an anomaly in the global symmetry or through explicit breaking terms.
 
\subsection{Anomaly}

If the global symmetry is anomalous then the triangle diagram generates a $a G \tilde{G}$ term which fits into a superpotential term
\begin{equation}
W_{\text{anomaly}}=-\frac{c_\text{an}}{f} A W^a W^a
\end{equation}
where $W^a=\lambda^a-i\sigma^{\mu\nu}\theta G^a_{\mu\nu}+\ldots$ is the field strength chiral superfield for the gauge group G which has a U(1)G$^2$ anomaly.  In practice we will take G to be SU(3)$_\text{color}$ or U(1)$_\text{QED}$ since we will be interested in the coupling to massless gauge fields. $W_\text{anomaly}$ generates non-derivative couplings in the effective Lagrangian:
\begin{equation}
\label{anomaly_int}
\mathcal{L}_{\text{anomaly}} \supset\frac{c_{\text{an}}}{f\sqrt{2}}
\left(a G^a_{\mu\nu}\tilde{G}^a_{\mu\nu}+\frac{i}{2}\bar{\chi }G^a_{\mu\nu}[\gamma^\mu,\gamma^\nu]\gamma^5\lambda^a \right)
\end{equation}
where $\tilde{G}^a_{\mu\nu}=\frac{1}{2}\epsilon_{\mu\nu\rho\sigma}G^a_{\rho\sigma}$. 

For the remainder of this document we assume that the global U(1) is anomalous.
For example, the anomalous coupling  $c_\text{an}$ is generated when the Goldstone boson $a$ couples to $N_{\psi}$ fermions $\Psi_i$ that transform in the fundamental of the gauged SU($N$)
and carry a global charge $q_{\Psi}$,
\begin{equation}
c_{\text{an}}=\frac{\alpha}{8\pi} \sqrt{2}\sum^{N_\Psi}_i \left(\frac{y_i f}{m_{\Psi_i}}\right)\qquad\qquad\qquad 
\mathcal{L}_{y}=ia\sum_{i=1}^{N_{\Psi}} y_i \bar{\Psi}_i\gamma^5\Psi_i \,.
\end{equation} 
The result for a U(1) gauge group is similar and is obtained by including the different $q_i$ charges,
\begin{equation}
c^{(1)}_\text{an}=\frac{\alpha_{U(1)}}{8\pi} \sqrt{2} N_c \sum_i 2q_i^2 \left(\frac{y_i f}{m_{\Psi_i}}\right),
\end{equation}
where $N_c=3$ and the factor of $2$ comes from the normalization of the generators in SU($N$), Tr$[T^a T^b]=\delta^{ab}/2$.
The simplest and most common case is when all masses are degenerate, $m_{\Psi_i}=m_{\Psi}$, and the $y_i$ are equal,  $y_i=m_{\Psi} q_{\Psi}/(f\sqrt{2})$, so that 
\begin{equation}
c_{\text{an}}=\frac{\alpha}{8\pi} q_{\Psi} N_{\Psi}\,.
\end{equation}

Note that gauge coupling unification is preserved if the mediator fields $\Psi_i$ are embedded in complete GUT multiplets.
For example, one may consider $N_{\Psi}\times(\mathbf{5}\oplus\bar{\mathbf{5}})$ representations of SU(5) which decompose into\footnote{The hypercharge normalization is fixed if there are no exotic electric charges.} $(3,1)_{1/3}$ and $(1,2)_{-1/2}$ under $\text{SU}_{\text{c}}(3)\times \text{SU}_\text{L}(2)\times \text{U}_{Y}(1)$. In this case the mediators $\Psi_i$ are both colored and electrically charged; they thus allow the dominant decay to be $a\rightarrow gg$ with subdominant contributions from $a\rightarrow \gamma\gamma$ with branching ratio $\sim 10^{-3}$.

\subsection{Explicit breaking spurions}
 
Sources of explicit global symmetry breaking terms can be parametrized by spurion chiral superfields $R_{\alpha}$ which carry charge $\alpha$ under the global symmetry and obtain a vev  $\langle R_{\alpha} \rangle = \lambda_\alpha f$, where $\lambda_\alpha \ll 1$. This permits a superpotential term $\Delta W= f^2 \sum_\alpha R_{-\alpha} G^\alpha$ where $G^\alpha=\exp(\alpha A/f)$. 
Unbroken supersymmetry requires that there are two sources of global symmetry breaking, $R_{-\alpha}$ and $R_{-\beta}$, with opposite charges, $\alpha\beta<0$.
This produces a sGoldstone boson vev and an effective superpotential with a common supersymmetry preserving masses  $m_{\chi }=m_{a}=m_{s}$. Explicit breaking terms may also generate new interactions which are completely determined by the Goldstone boson mass,
\begin{equation}
\label{new_interactions}
\mathcal{L}\supset 
-\frac{m_a^2}{2}\left(a^2+s^2\right)-\frac{m_a}{2}\bar{\chi}{\chi}+
\frac{m_a}{2\sqrt{2}f}(\alpha+\beta)\,\left(i a \bar{\chi }\gamma^5\chi-s \bar{\chi }\chi-m_a s a^2\right)+\frac{m_a }{8 f^2}(\alpha^2+\alpha\beta+\beta^2)\, a^2\bar{\chi }\chi +\ldots
\end{equation}
The only model-dependent inputs are the charges $\alpha$ and $\beta$ of the explicit breaking operators.
After SUSY breaking, the sGoldstone boson and the Goldstone fermion masses are lifted  while the Goldstone boson remains light, $m_{s}\gg m_{\chi} > m_{a}$.
Up to integration by parts, the the on-shell trilinear axial coupling $a\bar{\chi}\gamma^5\chi$ is equivalent to an effective $b_1$ coupling in (\ref{Lagrangian_self}).
Finally, the explicit breaking can also generate additional terms in the superpotential of the form 
\begin{equation}
W=  c_i R_{-\alpha} G^\alpha  \left( H_u H_d + \frac{Y_u}{M_u} \bar{Q} H_u u +\frac{Y_d}{M_d} \bar{Q} H_d d +\frac{Y_l}{M_L} \bar{L} H_d e +\text{h.c.} \right)\,.
\end{equation}
These lead to mixing with the Higgs and decays to SM fermions.

\section{Relic Abundance}
\label{Abundance_sect}

The Goldstone fermion $\chi$ is a natural dark matter candidate if it is the LSP and  produces the observed abundance \cite{Larson:2010gs,Nakamura:2010zzi},
\begin{equation}
\Omega_{\text{DM}} h^2=0.112\pm 0.0056,
\end{equation}
where $h$ is the Hubble constant. 
A key observation is that the effective interactions between the Goldstone fermion $\chi$ and Goldstone boson $a$ lead to an annihilation cross section $\chi\chi\rightarrow aa$ of the correct magnitude for a thermal relic with $\mathcal{O}(1)$ couplings and mass at the SUSY breaking scale $M_{\text{SUSY}}$,
\begin{align}
\label{annihilation_approx}
\langle\sigma v \rangle &\approx  \frac{b_1^4 }{8\pi}\frac{T_f}{m_{\chi}}\frac{m_{\chi}^2}{f^4}  \simeq 1\mbox{ pb}\\
 \Omega_{\text{DM}} h^2 &\approx  \frac{0.1\text{ pb}}{\langle \sigma v\rangle}\,.
\end{align}
Note that an explicit factor of the temperature appears in (\ref{annihilation_approx}) because parity forbids the $s$-wave channel. Thus the Goldstone fermion is an almost ideal WIMP candidate. Due to the slight thermal suppression, the coupling $b_1$ has to be slightly larger than 1. Otherwise, with the natural choices of parameters, the correct annihilation cross section is obtained. 

\subsection{Summary of model parameters}
\label{sec:summary:of:parameters}

Below we provide a summary of the Goldstone fermion model parameters and the values used in our parameter space scan:
	\begin{center}
	\begin{tabular}{|c|l|l|}
		\hline
		\textbf{Parameter} & \textbf{Description} & \textbf{Scan Range}\\
		\hline
		$f$ & Global symmetry breaking scale & $500\text{ GeV} - 1.2\text{ TeV}$\\
		% \hline
		$m_\chi$ & Goldstone fermion mass & $50 - 150$ GeV\\
		% \hline
		$m_a$ & Goldstone boson mass & $8$ GeV -- $f/10$\\
		% \hline
		$b_1$ & $\chi\chi a$ coupling, (\ref{eq:b1:expression}) & $[0,2]$\\
		% \hline
		$c_\text{an}$ & Anomaly coefficient, (\ref{anomaly_int}) & $0.06$\\
		% \hline
		$c_h$ & Higgs coupling, (\ref{Higgs_coupling}) & $[-1,1]$\\
		% \hline
		$\delta = (\beta-\alpha)/2$ & Explicit breaking $ia\bar\chi\gamma^5\chi$ coupling, (\ref{new_interactions}) & $3/2$\\
		% \hline
		$\rho = (\alpha^2+\alpha\beta+\beta^2)/8$ & Explicit breaking $a^2\bar\chi\chi$ coupling, (\ref{new_interactions}) & $13/8$\\
		\hline
	\end{tabular}
	\end{center}
These values represent a natural cross section of the full parameter space.  
%The $b_1$, $\delta$, and $\rho$ parameters may also take negative values.

\subsection{Summary of annihilation channels}

\begin{figure}[htb]
\begin{center}
\includegraphics[scale=0.7]{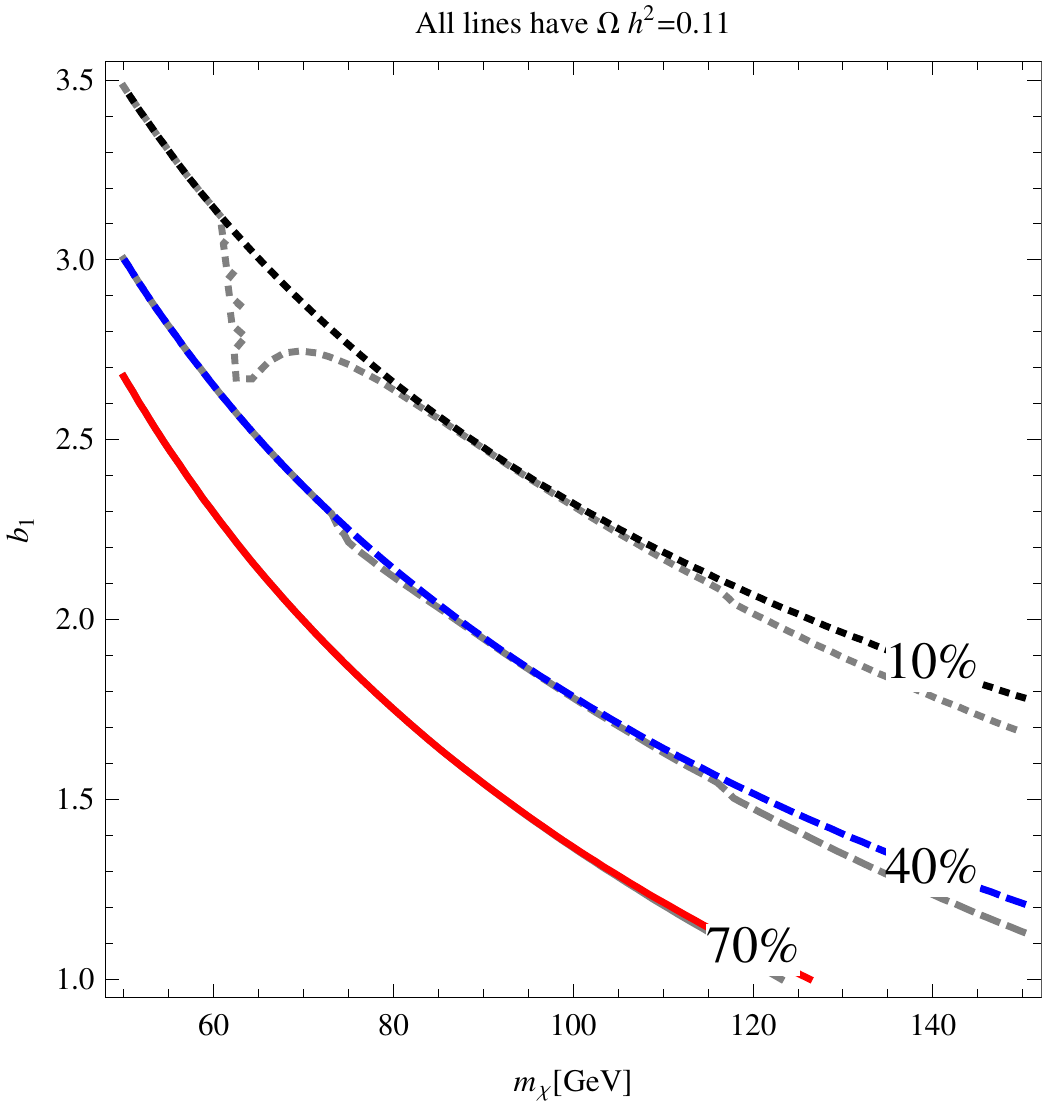}
\caption{
Contours for different values of the Goldstone boson mass: $m_a/m_\chi=0.1$~(black dotted), $0.5$~(blue dashed), and $0.7$~(red solid) for fixed relic density $\Omega h^2=0.11$ in the $(m_{\chi}, b_{1})$ plane.
% Contours of fixed relic density $\Omega h^2=0.11$ in the $(m_{\chi}, b_{1})$ plane for different values of the Goldstone boson mass: $m_a/m_\chi=0.1$~(black dotted), $0.5$~(blue dashed) and $0.7$~(red solid).
%
Gray lines include the subleading contributions from annihilations into Higgs bosons, $\chi\chi\rightarrow ah$ and $\chi\chi\rightarrow hh$. The kink at 60 GeV comes from threshold effects. We set $f=700$~GeV, $\alpha=-4 $, $\beta=1$, $m_h=116$~GeV, $c_\text{an}=0.06$.
}
\label{fig_abundance}
\end{center}
\end{figure}

In addition to
\begin{itemize}\addtolength{\itemsep}{-0.75\baselineskip}
\item[(a)] $\chi \chi \rightarrow aa$ in the $t$-channel and $u$-channel via the self-interactions (\ref{Lagrangian_self}),
\end{itemize}
a detailed analysis shows that there may be appreciable contributions from 
\begin{itemize}\addtolength{\itemsep}{-0.5\baselineskip}
%\item[(a)] $\chi \chi \rightarrow aa$ via a t-channel and u-channel; 
\item[(b)] $\chi \chi \rightarrow aa$ from explicit breaking terms (\ref{new_interactions});
\item[(c)] $\chi \chi \rightarrow gg$ with $a$ in the $s$-channel via the anomaly (\ref{anomaly_int}).
\end{itemize}
In fact, these can overcome the $p$-wave suppression in the annihilation into $2a$.  Note that $\chi \chi \rightarrow gg$ also gives an $s$-wave contribution which may contribute up to $\sim1/3$ of the total annihilation cross section. Less significant are the decays into Higgs bosons,
\begin{itemize}\addtolength{\itemsep}{-0.5\baselineskip}
\item[(d)]$\chi\chi\rightarrow a h$   with $a$ in the $s$-channel via the Higgs coupling (\ref{Higgs_coupling}) when $m_h+m_a<2m_\chi$;
\item[(e)] $\chi \chi \rightarrow  hh$ via the coupling to two Higgs bosons (\ref{generic_Higgs}) when $m_{h}< m_\chi$.
\end{itemize}
Note that in some cases the Higgs boson may be lighter than the 115 GeV because of non-standard Higgs decays (see Section \ref{colliderPheno} for the relevant collider phenomenology). Other annihilations involving a virtual gluino, the sGoldstone, or the Higgs boson are typically suppressed by large masses or small Yukawa couplings. A detailed calculation of each contribution is presented in Appendix \ref{appAnnihilation}. The model generates the correct abundance for Goldstone fermion masses between $50 - 150$ GeV and Goldstone boson masses between $10\% - 100\%$ of $m_\chi$ for couplings $b_1\sim \mathcal O(1)$. Fig.~\ref{fig_abundance} shows the contours for different values of $m_a/m_{\chi}$ subject to the correct relic abundance in the $(m_{\chi}, b_1)$ plane.
It may further be possible to open up a different region of parameter space with lighter Goldstone boson and fermion scales through a Sommerfeld enhancement due to an attractive force between the exchange of multiple low-energy Goldstone bosons \cite{Bedaque}. We briefly discuss this possibility in Appendix~\ref{app:Sommerfeld}

\section{Direct and Indirect Detection}
\label{sec:direct_indirect_detection}

We have seen above that this model can easily produce the correct dark matter abundance. Next we estimate the generic size of direct and indirect detection bounds.

\subsection{Dark Matter Effective Operators for Direct detection}

In order to evaluate the cross section of Goldstone fermion scattering off nuclei in direct detection experiments, one must evaluate the nucleon matrix elements.
Usually one parametrizes the light quark mass content  $f^{(N)}_i$ of the nucleons,
\begin{equation}
\label{f1}
m_{i}\langle N|\bar{q}_i q_i |N\rangle\equiv f^{(N)}_{i} m_{N}
\qquad\qquad\qquad
i=u,d,s\,,
\qquad\qquad\qquad
N=p,n.
\end{equation}
where $m_{N}$ is the nucleon mass. The largest contribution comes from the strange quark \cite{Cheng:1988cz}, but with sizeable uncertainties \cite{Ellis:2008hf}. We assume the default value in the micrOMEGAs code, 
$f^{(N)}_{u,d}\ll f^{(p)}_s=f^{(n)}_s=0.26$ \cite{Belanger:2008sj}. For heavy quarks, the contribution $f^{(N)}_h$ is induced via gluon exchange and can be calculated by means of the conformal anomaly \cite{Shifman:1978zn},
\begin{equation}
\label{f2}
m_{h}\langle N| \bar{q}_h q_h |N \rangle\equiv f^{(N)}_{h} m_{N}=
\frac{2}{27}m_{N}\left(1-\sum_{i=u,d,s}f^{(N)}_i\right)\,,
\qquad\qquad
 h=c,b,t\,.
\end{equation}

\subsubsection{Coupling to quarks via  Higgs exchange}

\begin{figure}[htb]
\begin{center}
\includegraphics[scale=0.8]{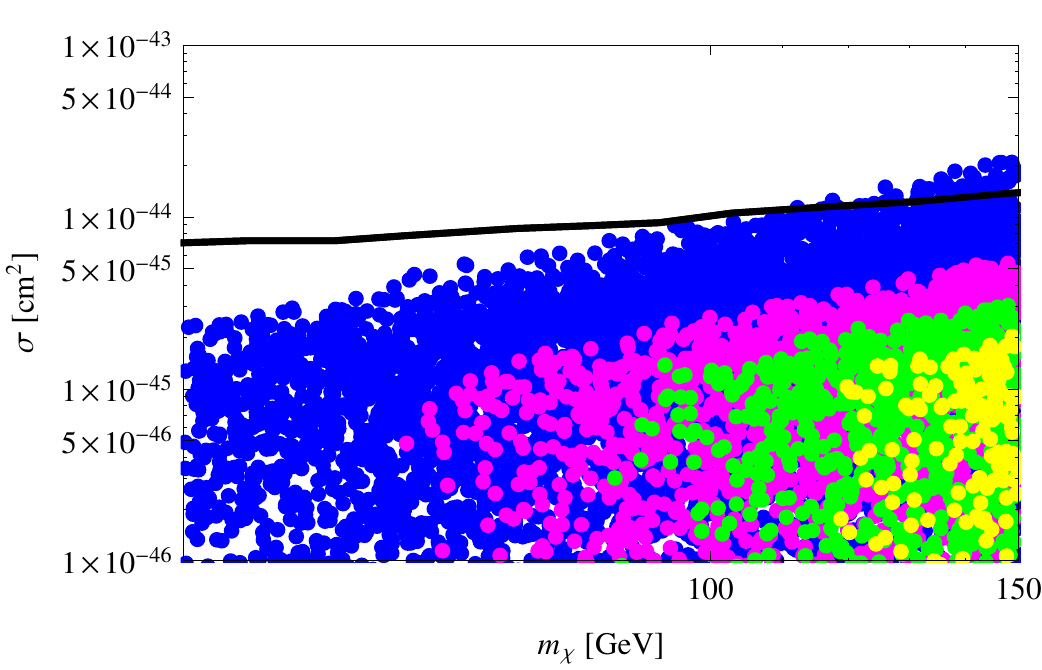}
\includegraphics[scale=0.8]{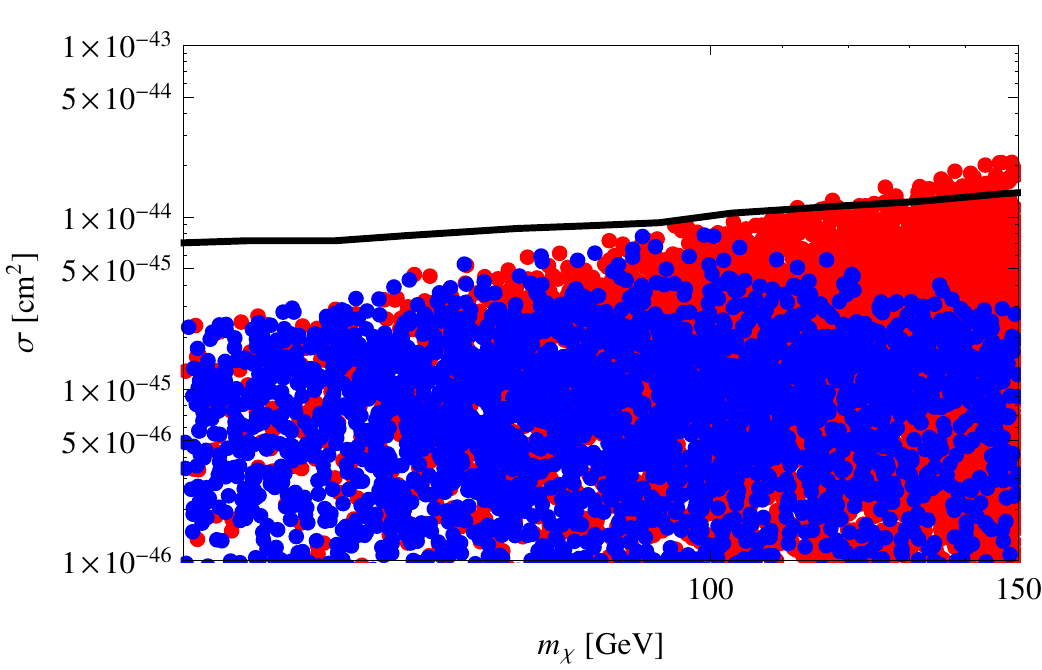}
 \caption{Black line: XENON100 bound. Left: scan over parameter space with $500<f<700$~GeV (blue), $700<f<800$~GeV (violet), $800<f<900$~GeV (green), $900<f<1000$~GeV (yellow). We scan $0<b_1<2$, $-1<c_{h,hh}<1$, $8\mbox{ GeV}\,<m_a<\frac{f}{10}$, $50\mbox{ GeV}\,<m_{\chi}<200$~GeV.  Right:  Blue points have $0.5<m_{a}/m_{\chi}$ whereas red points have $m_{a}/m_{\chi}<0.5$.
}
\label{fig:paramvsxenon}
\end{center}
\end{figure}

In Section~\ref{interactions_lightFields} we showed that after electroweak symmetry breaking, it is natural to expect a non-vanishing coupling between $\chi$ and the lightest neutral Higgs boson $h$,
\begin{equation}
\mathcal{L}_{h}=c_h \frac{v_\text{EW}}{2f^2}(\bar{\chi} i\gamma^\mu \partial_\mu\chi)h\,,
\end{equation}
where the size of the coupling $c_{h}$ depends on the specific realization. The Higgs coupling to nucleons is set by the Yukawa couplings and---in the presence of more Higgses---the mixing angles, $c_{q} m_{q}/(\sqrt{2}v_\text{EW}) h\bar{q}q$.
Integrating out the Higgs generates an effective four-Fermi interaction,
\begin{equation}
\mathcal{L}^\text{eff}_{\chi N} =G_{\chi N}\bar{N}N\bar{\chi}\chi
\qquad\qquad\qquad
G_{\chi N}=c_{h}\frac{\lambda_{N}}{2\sqrt{2}}\left(\frac{m_{\chi}m_{N}}{m_{h}^2 f^2}\right)\, ,
\end{equation}
where we used the equations of motion for $\chi$ and the quark content of the nucleons (\ref{f2}) to write
\begin{equation}
\lambda_{N}=\sum_{q=u,d,s}c_q f^{(N)}_{q}+\frac{2}{27}\left(1-\sum_{q=u,d,s}f^{(N)}_{q}\right)  \left(\sum_{q^\prime=c,b,t}c_{q^\prime}\right)\,.
\end{equation}
The resulting scattering cross section per nucleon at zero momentum transfer is\footnote{
Since most direct detection events occur at low recoil energy, it is standard to parameterize the cross section in terms of a zero momentum transfer part and a form factor which encodes the momentum and target dependence. See, for example, \cite{Engel:1991wq}.
}
\begin{align}
\sigma_\text{SI}^\text{Higgs}= \frac{4\mu_\chi^2}{A^2 \pi} \left[G_{\chi\,p} Z+G_{\chi\,n} (A-Z)\right]^2\,,
\end{align}
where $\mu_{\chi}=(m_{\chi}^{-1}+m_{N}^{-1})^{-1}$ is the reduced mass. 
The typical value for $\sigma_\text{SI}^\text{Higgs}$ is just below the XENON100 direct detection bound \cite{Aprile:2011hi},
\begin{align}
\sigma_\text{SI}^\text{Higgs}
\approx  3 c_{h}^2 \times 10^{-45}\mbox{ cm}^2\,
\left(\frac{115\mbox{ GeV}}{m_{h}}\right)^4
\left(\frac{700\mbox{ GeV}}{f}\right)^4
\left(\frac{m_{\chi}}{100\mbox{ GeV}}\right)^2 
\left(\frac{\mu_\chi}{1\mbox{ GeV}}\right)^2
\left(\frac{\lambda_N}{0.5}\right)^2\, .
\end{align}
Note the $(m_\chi v/f^2)^2$ suppression present in this cross section (due to the Goldstone nature of $\chi$) relative to that of a generic Higgs exchange.
For example, Higgs-mediated neutralino decay in the MSSM with coupling $\mathcal{L}\approx cg/2 \bar{\chi}\chi h$ needs a very small coupling $c$ to avoid the XENON100 bounds:
\begin{equation}
\sigma^\text{MSSM}_\text{SI}\sim \frac{c^2 g^2}{2\pi}\frac{\lambda_{N}^2 \mu^2 m_{N}^2}{m_{h}^4 v_\text{EW}^2}\approx c^2\times 10^{-42}\mbox{ cm}^2\,.
\end{equation}
Thus, Goldstone fermion dark matter offers a natural suppression of the direct detection cross section while retaining the correct WIMP annihilation cross section and abundance.
Fig.~\ref{fig:paramvsxenon} plots typical values of the direct detection cross section for parameters with correct relic abundance.

\subsubsection{Coupling to gluons}

Integrating out the massive gaugino in (\ref{anomaly_int}) generates two dimension-$7$ operators that couple $\chi $ to gauge bosons,
\begin{align}
\mathcal{L}_\text{eff}^{(1)}=-\left(\frac{c^2_\text{an}}{2M_{\lambda} f^2}\right)\left[\bar{\chi }\chi \right]G^a_{\alpha\beta}G^a_{\alpha\beta}
\qquad \qquad \qquad
\mathcal{L}_\text{eff}^{(2)}=-i \left(\frac{c^2_\text{an}}{2M_{\lambda} f^2}\right)\left[\bar{\chi }\gamma^5\chi \right]G^a_{\alpha\beta}\tilde{G}^a_{\alpha\beta}\, ,\label{gluon-operators}
\end{align}
where $M_{\lambda}$ is the gaugino mass. In the limit of zero momentum transfer,
only $\mathcal{L}_\text{eff}^{(1)}$ contributes to direct detection since $G\tilde{G}$ is a total derivative. We therefore neglect $\mathcal{L}_{eff}^{(2)}$ hereafter.
The $\langle N|GG|N\rangle$ nucleon matrix element can be extracted 
from the conformal anomaly (\ref{f2}) so that  $\mathcal{L}_\text{eff}^{(1)}$ can be mapped to a standard four-Fermi operator
\begin{equation}
\label{effective_4fermi}
 \mathcal{L}^{(1)}_\text{eff} \longrightarrow  \mathcal{L}^{(1)}_{\text{eff},\, N}=G_N \bar{\chi }\chi  \bar{N}N\,,
\qquad G_N=\frac{4\pi c_\text{an}^2}{9\alpha_s}\frac{m_{N}}{M_{\lambda} f^2} \left(1-\sum_{i=u,d,s}f^{(N)}_i\right)\,.
% \ \,,\qquad  N=p,n
\end{equation}
The corresponding cross section per nucleon at zero momentum transfer is
\begin{align}
\sigma_\text{SI}^{gg}
\approx 2\times 10^{-48}\mbox{ cm}^2\,
\left(\frac{700\mbox{ GeV}}{M_{\lambda}}\right)^2 
\left(\frac{700\mbox{ GeV}}{f}\right)^4 \left(\frac{N_{\Psi}}{5}\right)^4\left(\frac{q_{\Psi}}{2}\right)^4 \left(\frac{\mu}{1\mbox{ GeV}}\right)^2
\end{align}
where $c_\text{an}=\alpha_s q_{\Psi} N_{\Psi}/(8\pi)$ has been used.
This value is much smaller than both the recent upper bound by the XENON100 collaboration \cite{Aprile:2011hi} and the expected reach at the LHC, 
$\sigma^\text{SI}_{gg}=\mbox{few}\times 10^{-46}\mbox{ cm}^2$ \cite{Goodman:2010yf}.

\subsection{Indirect detection}

Many experiments are searching for indirect signals of annihilation of dark matter in dense environments such as the galactic center or the solar core. The rate of such events is set by the present-day thermally averaged annihilation cross section.
Note, however, that the dominant annihilation channels at freeze-out are $p$-wave and hence are strongly velocity suppressed in the current era. 
Thus, the relevant annihilation channels for indirect detection are $s$-wave and were sub-dominant at freeze-out.  These cross sections are relatively small and astrophysical observations do not impose severe constraints.

\subsubsection{Fermi-LAT: lines, isotropic diffuse $\gamma$-rays, and dwarf galaxies}

Dark annihilation in the galactic halo may produce photons either
directly (e.g. $\chi\chi \rightarrow \gamma \gamma$) or through secondaries 
(bremsstrahlung off charged products or decays of neutral pions).
The Fermi experiment has searched for excesses in the gamma ray spectrum both in the form of lines arising from prompt annihilation to photons and in contributions to the diffuse spectrum from secondary products of annihilation.

Fermi currently searches for $\gamma$-ray lines from 30 -- 200 GeV \cite{Abdo:2010nc}, with upcoming bounds that are an order of magnitude stronger in the 7 -- 30 GeV region \cite{private}.
The lack of a bump in the Fermi data implies an upper bound on $\langle \sigma v \rangle_{\gamma\gamma}$ between $(0.2 - 2.5) \cdot 10^{-27}
\text{cm}^3/\text{s}$ when using the Einasto dark matter halo profile which predicts the largest photon flux among those examined in the Fermi analysis.

In the Goldstone fermion model, prompt annihilation to photons occurs through an anomaly vertex similar to the one which mediates $a \rightarrow gg$.  This rate depends on the $\text{U}(1) \times \text{U}(1)_\text{EM}^2$ anomaly coefficient which is determined by the choice of electric charges for fields carrying global charge.  
The cross section for annihilation into gluons is given in (\ref{eq:ggann}).  The analagous expression for annihilations into photons is given by replacing $\alpha_s^2 N_c \rightarrow  \alpha_\text{EM}^2\left( 2 \sum_i (q_\text{EM}^i)^2 \right)^2$.  For the case where the $\Psi$ are taken to be in the (anti-)fundamental of an $\text{SU}(5)$ unified group, we find $\langle \sigma v \rangle_{\gamma\gamma} \sim 2 \cdot 10^{-3} \langle \sigma v \rangle_{gg}$.  Even with the most extreme choices of the model's free parameters, this rate remains more than an order of magnitude smaller than the Fermi bounds.

Fermi has also measured the isotropic diffuse $\gamma$-ray spectrum in the range  $20-100$ GeV  \cite{Abdo:2010dk}.  
This bounds the annihilation of dark matter into charged particles and neutral pions.  For example, for a 400 GeV dark matter particle which annihilates into a $b\bar{b}$ pair, Fermi sets a bound on $\langle \sigma v\rangle_{b \bar{b}}$ which is roughly an order of magnitude above the cross section required to reproduce the right relic abundance. 
The Goldstone fermion model generates diffuse photons primarily through annihilation to gluons produced in the $s$-wave annihilation channel $\chi \chi \rightarrow gg$.  
However, this cross section is at least an order of magnitude smaller than the Fermi bound and hence the Fermi diffuse $\gamma$-ray data do not constrain this model.

Preliminary results from a Fermi analysis of 10 dark-matter-rich dwarf
spheroidal galaxies also place limits on photo-production from dark matter
annihilation~\cite{moreprivate}.  For low-mass ($\lesssim 60$ GeV) dark matter
annihilating into $b \bar{b}$ pairs, constraints on the annihilation
rate extend slightly below the thermal relic rate of 
$3 \cdot 10^{-26}~\text{cm}^3/\text{s}$, 
with the strongest constraint of $\sim 1 \cdot 10^{-26} ~\text{cm}^3/\text{s}$ at $m_{\chi} = 10$~GeV.  In
this mass window and for reasonable parameter choices, the Goldstone fermion annihilation cross section is always at least a factor of $3$
lower than these limits.

Other constraints, such as those that come from $\gamma$-rays originating in clusters of galaxies, typically set weaker bounds \cite{Ackermann:2010rg}.

\subsubsection{PAMELA: the antiproton flux}

\begin{figure}[htb]
\begin{center}
\includegraphics[scale=.75]{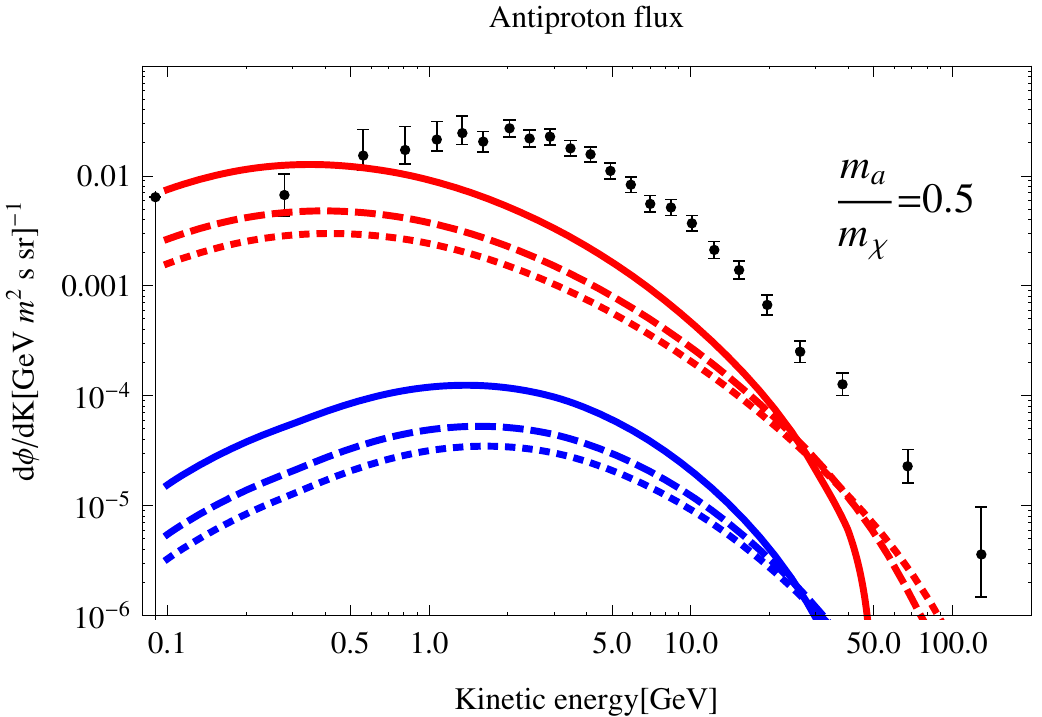}
\includegraphics[scale=.75]{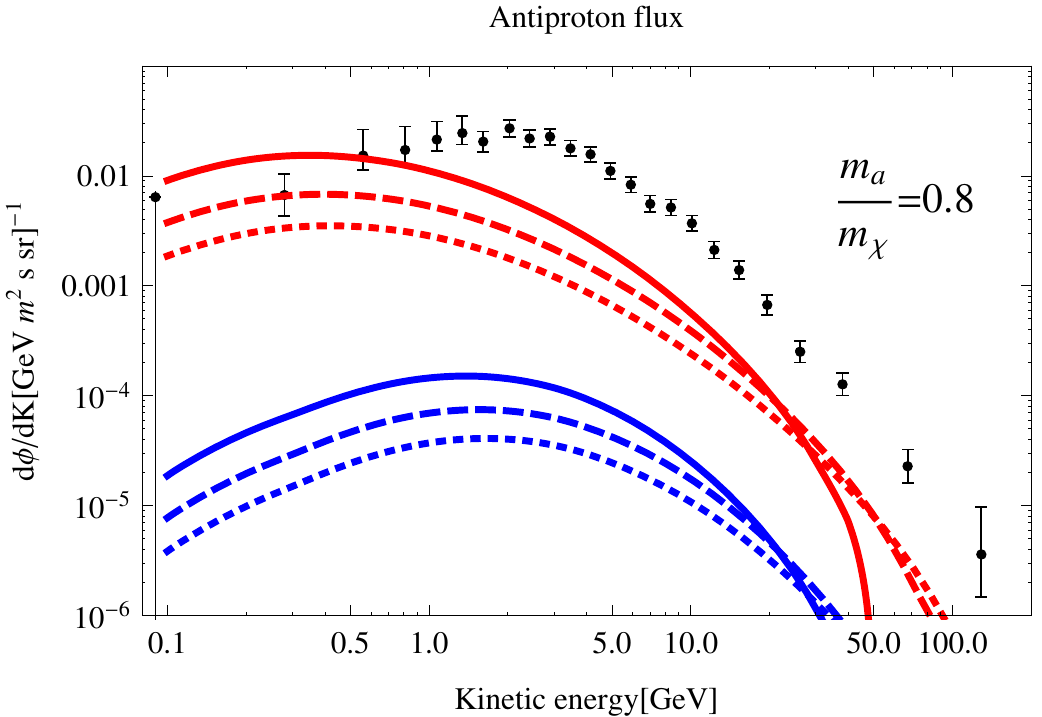}
 \caption{
Antiproton flux at the Earth for $f=700$~GeV, $Q_{\Psi}=2$, $\delta=3/2$, $N_{\Psi}=5$ at fixed density $\Omega h^2\simeq 0.1$.  
Red (blue) lines represent the propagation parameters MAX (MIN) used in \cite{Cirelli:2010xx} with the Einasto DM halo profile.
The dots represent the PAMELA data \cite{Adriani:2010rc}. 
 Left: $m_a/m_\chi=0.5$ at $m_\chi=50$ GeV and $b_1=3$ (solid); $m_\chi=100$ GeV and $b_{1}=1.5$ (dashed); $m_\chi=150$ GeV and $b_1=1 $(dotted). 
 Right: $m_a/m_\chi=0.8$ at $m_\chi=50$ GeV and $b_1=2.5$ (solid); $m_\chi=100$ GeV and $b_{1}=1.2$ (dashed); $m_\chi=150$ GeV and $b_1=0.5 $(dotted).  }
\label{fig:Antiprotonfluxes}
\end{center}
\end{figure}

PAMELA has recently published data on the absolute cosmic ray antiproton flux from $60$ MeV -- $180$ GeV  \cite{Adriani:2010rc}. 
This places constraints on dark matter models with a substantial annihilation rate to hadrons.  
For a 100 GeV WIMP, the annihilation cross section to $Z$s, $W$s, and $b$ quarks has an upper bound comparable to the rate required for the observed relic abundance, $\langle \sigma v \rangle_\text{relic} \sim 3 \cdot 10^{-26} \text{cm}^2/\text{s}$~\cite{Cholis:2010xb}.
For Goldstone fermion dark matter, the dominant annihilation channel in the galactic halo, $\chi\chi\to gg$, is $s$-wave. This has a typical cross section of $\langle\sigma v\rangle\sim 10^{-27}\mbox{ cm}^3/\mbox{s}$ and can be pushed up as high as $10^{-26}\mbox{ cm}^3/\mbox{s}$.
Using recent numerical recipes \cite{Cirelli:2010xx}, one may estimate the anti-proton flux as a function of the thermally averaged annihilation cross section and the Goldstone fermion mass. This is depicted in Fig.~\ref{fig:Antiprotonfluxes} for different model parameters that yield the correct relic abundance.
The anti-proton flux varies considerably as a function of the galactic propagation parameters and the halo profile. The solid, dashed, and dotted curves each correspond to different underlying Goldstone fermion model parameters.  Choosing different halo profiles and propagation parameters leads to a spread in the predicted $\bar{p}$ flux such that for each choice of Goldstone fermion model parameters, the actual flux from dark matter annihilation is expected to lie between the two solid, dashed, or dotted curves respectively.

For each choice of model parameters, there is a sizeable region where the predicted flux from dark matter annihilation lies well below the measured anti-proton flux.  Thus the PAMELA data do not place significant constraints on the Goldstone fermion dark matter model.

\section{Collider Phenomenology}
\label{colliderPheno}

In addition to (in)direct detection, Goldstone fermion models lend themselves to novel collider signatures coming from the Goldstone supermultiplet. As discussed in Section~\ref{susy_breaking_sect}, the sGoldstone $s$ is typically heavy with small couplings to the SM sector so we may neglect its collider signatures.

\subsection{Collider signals of dark matter}

The most direct way of testing the dark matter annihilation mechanism is through dark matter pair production coming, for example, from the $\bar \chi\chi GG$ operator in (\ref{gluon-operators}).
One signature of this process at colliders is a monojet coming from hard initial state QCD radiation.
For a range of masses up to the TeV scale, the LHC will set the most stringent bound on this operator with a sensitivity of $\sigma_{\text{SI}}^{N}\sim 10^{-46} -10^{-45} \text{cm}^2$ for a $5\sigma$ discovery with 100 fb$^{-1}$ \cite{Goodman:2010yf}.
The effective scale that suppresses the dimension-7 operator (\ref{gluon-operators}) is roughly an order of magnitude larger than the LHC reach. However, the process $gg\rightarrow a^{*} \rightarrow \chi\chi$ via an off-shell Goldstone boson gives a larger contribution and results in a naive effective scale $M^{*} \sim (m_{\chi} f^2/c_\text{an})^{1/3}\sim 1$~TeV. This is in the ballpark of the LHC $5\sigma$ reach given in \cite{Goodman:2010yf}. 

Goldstone fermion dark matter can also be produced from the cascade decay of heavier $R$-parity odd particles, such as gluinos or squarks. Due to the small coupling between the MSSM and the Goldstone sectors, the cascade decays will all go through the lightest ordinary supersymmetric particle (LOSP). The decay of LOSP to the Goldstone fermion is determined by the operators connecting the two sectors. In the current setup, there are two types of interactions:
\begin{itemize}\addtolength{\itemsep}{-0.5\baselineskip}
\item the anomaly induced coupling $\bar \chi G \lambda$, as in (\ref{anomaly_int}), and
\item the kinetic mixing discussed in Section~\ref{interactions_lightFields} .
\end{itemize}
The details of the decay modes depends on the nature of the LSP.
 For example, a bino-like LOSP will decay to the LSP via the anomaly, $\tilde{B}\rightarrow \chi+\gamma/Z$.
A Higgsino-like LOSP would decay instead to the LSP because of the kinetic mixing, $\tilde h \rightarrow \chi + h$, and  $\tilde h \rightarrow \chi + a\rightarrow \chi+2j$.  In the latter decay mode, the reconstruction of the Goldstone boson resonance in the jet final state is difficult if $a$ is below $100$~GeV, but it may be possible instead in the diphoton decays of $a$ with sufficient luminosity.
These channels yield prompt decays even though they may be suppressed by loops or small mixing angles. For example, the natural width for a pure bino LOSP is around $10^{-5}$~GeV. 

Finally, the presence of exotic heavy fermions $\Psi_i$ also has interesting implications at colliders. These fermions may be considered to be ``fourth generation'' quarks which, if they are sufficiently light, can be probed at the early stages of the LHC (see the discussion in \cite{Luty:2010vd} for an example).

\subsection{Non-standard Higgs boson decays}

\begin{figure}[htb]
\begin{center}
\includegraphics[scale=.75]{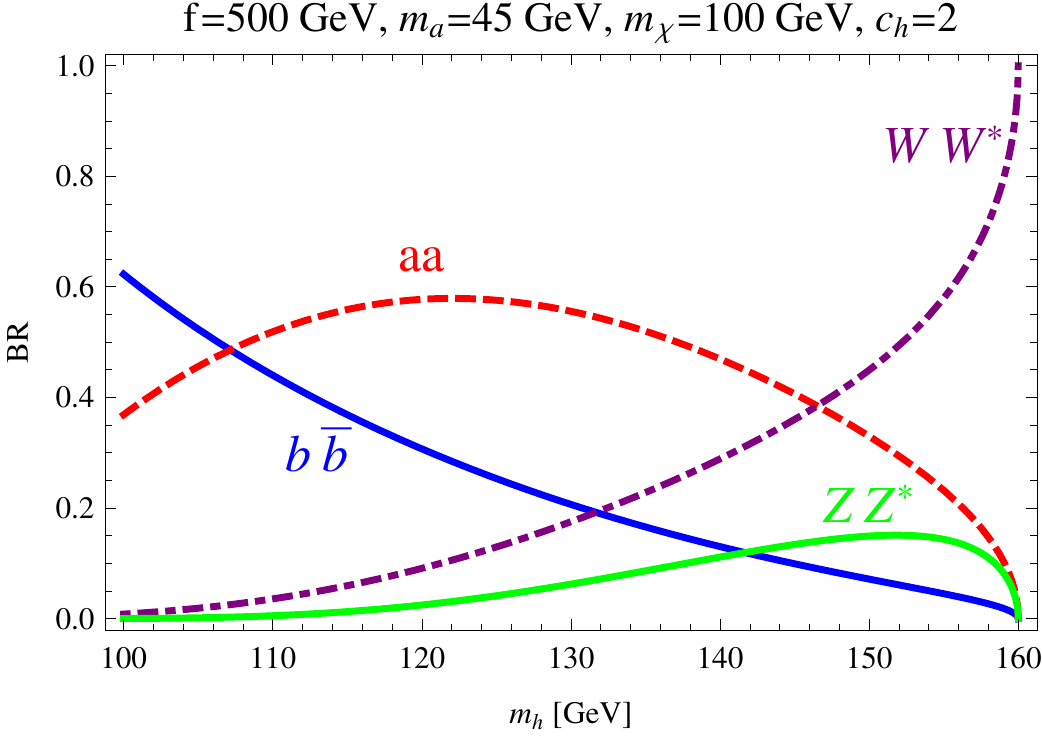}
\includegraphics[scale=.75]{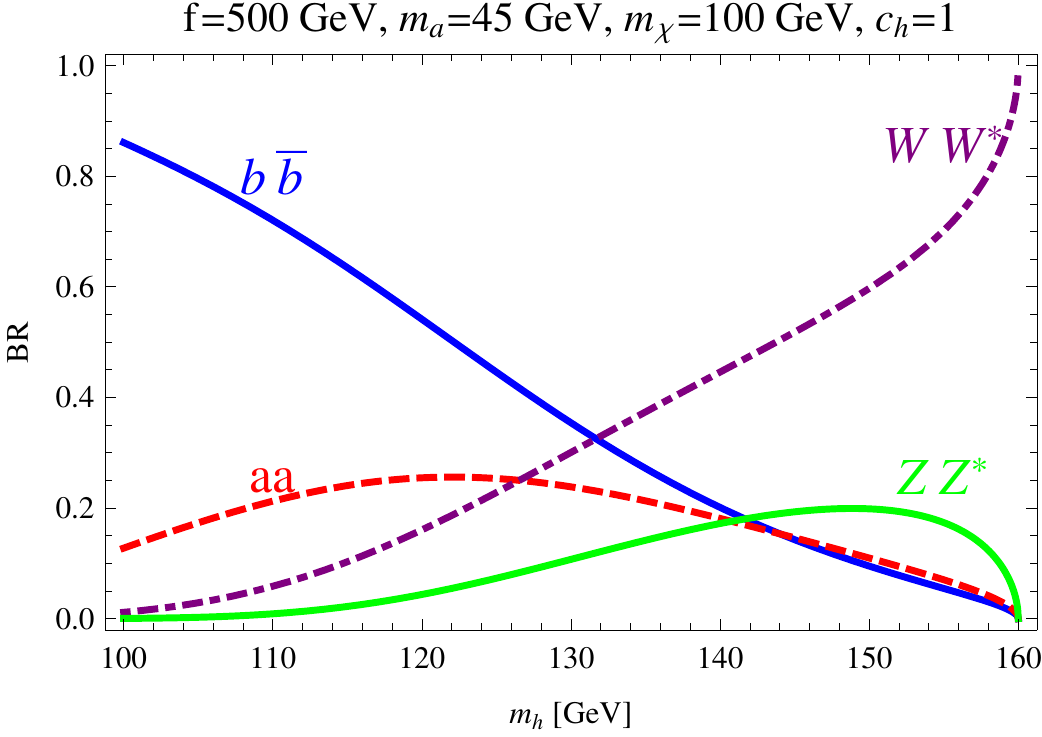}
% \caption{Higgs boson branching ratios.}
% \label{fig:br1plot}
% \end{center}
% \end{figure}
% %
% \begin{figure}[htb]
% \begin{center}
% %\includegraphics[scale=.75]{BR3.pdf}
\\
\includegraphics[scale=.75]{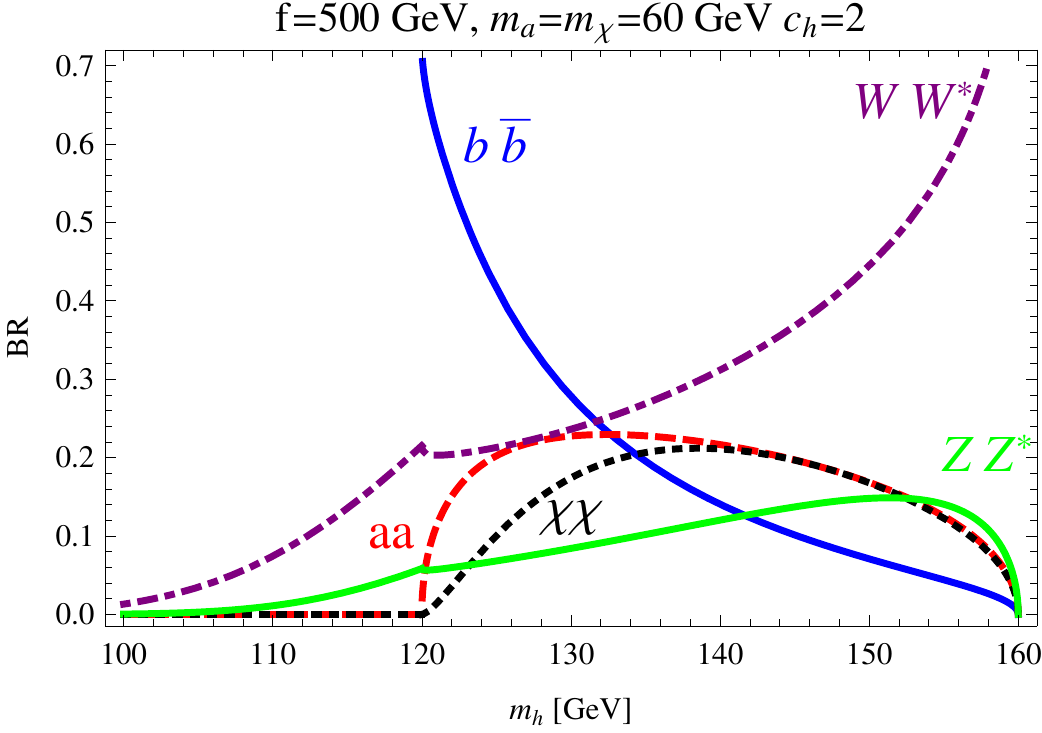}
\includegraphics[scale=.75]{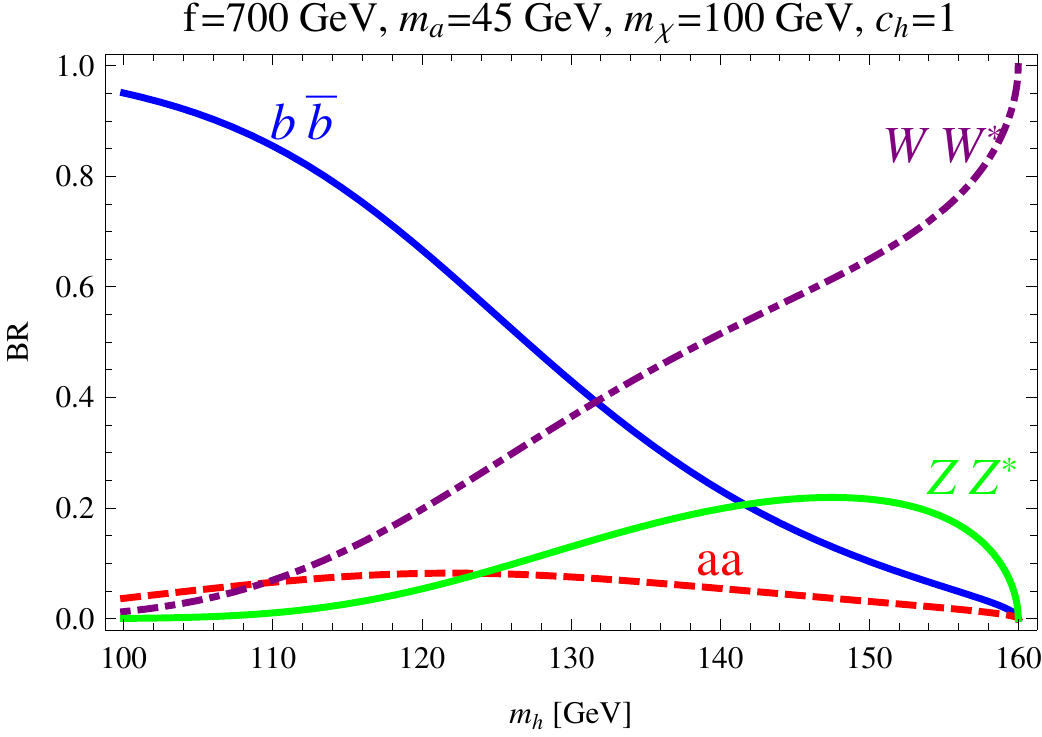}
\caption{Plots of Higgs boson branching ratios for various parameters.}
\label{fig:br2plot}
\end{center}
\end{figure}

The largest natural coupling of the Goldstone boson and fermion to the SM is through Higgs boson via the kinetic terms, (\ref{Higgs_coupling}). This coupling allows the Higgs to decay into $2a$ or $2\chi$ if kinematically allowed.  Typical branching ratios are plotted in Fig.~\ref{fig:br2plot}.

The Higgs boson decay $h\rightarrow 2a$ gives rise to four light, unflavored jets coming from $a\rightarrow 2g$. This decay mode is therefore easily `buried' under the QCD background. Such non-standard Higgs boson decays have recently been investigated in SUSY models where the Higgs boson itself is also a pseudo-Goldstone boson emerging  from the spontaneous breaking  of a global symmetry\footnote{For early attempts of this idea in SUSY see \cite{Dvali:1996cu}.  
More recently, SUSY and little Higgs models motivated by the little hierarchy problem have been proposed  \cite{Csaki:2005fc}.}. 
In particular, the spontaneous breaking of $\text{SU}(3)\rightarrow \text{SU}(2)$ gives rise both to a light Goldstone multiplet $A$ and a light Higgs multiplet \cite{Bellazzini:2009xt}. 
The resulting coupling $c_h\approx \sqrt{2}$ is set by the kinetic mixing between the two multiplets which, in turn, is fixed by the scale $f$ of the global symmetry breaking. 
A more recent example of a `buried Higgs' in SUSY has been discussed in the context of a spontaneously broken $\text{U}(1)$ symmetry where $c_h$ depends on couplings in the superpotential because the Higgs is no longer a pseudo-Goldstone boson \cite{Luty:2010vd}. 

Even though these non-standard Goldstone fermion decay modes can dominate, the branching ratio to SM particles is still larger than $\sim20\%$ at low Higgs masses and therefore the LEP bound on the Higgs mass cannot be lowered below $\sim110$~GeV.
Furthermore, while the discovery of a completely buried Higgs is challenging at the LHC \cite{Bellazzini:2010uk}, this `partially buried' Higgs would be discovered in SM channels with a missing piece in the total width. 
%but is still possible by means of jet substructure algorithms with sufficiently high luminosity \cite{Bellazzini:2010uk,Butterworth:2008iy}. 
% 
%
The invisible Higgs boson decays ($\chi$s leave the detectors) can be probed at the LHC through the missing energy signal \cite{Cavalli:2002vs}. 
Both the buried and invisible decay modes may have sizeable branching ratios, and the observation of both
channels would give strong evidence for this scenario.   

\section{Conclusions}

Acceptable dark matter scenarios within the MSSM must become increasingly contrived as the sensitivity of direct detection experiments increases.  In order to remain consistent with recent XENON100 results, neutralino WIMP models must typically invoke accidental mass relations to boost the annihilation cross-section through co-annihilations or strategically placed resonances.

Inspired by this tension, we have explored a general supersymmetric framework compatible with GUT unification in which the LSP is the fermionic component $\chi$ of a Goldstone supermultiplet associated with a U(1) global symmetry that is spontaneously broken at the TeV scale. 
Because the Goldstone fermion's couplings to the Standard Model are suppressed by $\sim v_\text{EW} m_{\chi}/f^2$ (and additional loop factors in some cases), these models are able to avoid direct detection constraints from XENON100 and indirect detection constraints from Fermi and PAMELA.

The annihilation cross section of a weak-scale Goldstone fermion at freeze out is on the order of 1 pb, with dominant contributions coming from $p$-wave annihilation into Goldstone bosons.   Typically subdominant $s$-wave annihilations into gluons arise through anomalies of the new global symmetry.  The observed dark matter relic density is obtained with natural values for the model parameters.

This class of models also offers novel and distinctive signatures at colliders. Goldstone fermions can be produced at the LHC in pairs through the anomalous coupling to gluons, leading to monojet signals when there is additional hard QCD radiation from the initial state.  Additionally, SUSY cascades are modified by decays of the NLSP to the Goldstone fermion. Examples include the bino decay to a photon and the Goldstone fermion, and the higgsino decay to the Goldstone fermion and the Goldstone boson.
The Goldstone multiplet also modifies the phenomenology of the Higgs sector. Interactions with the Goldstone boson allow cascade decays of the Higgs to four jets, $h\rightarrow 2a\rightarrow 4j$, analogous to models where the Higgs decays are `buried' under the QCD background. 
If kinematically allowed, the Higgs may also have a sizeable fraction of `invisible' decays, $h\rightarrow \chi\chi$. 

\subsection*{Acknowledgments}
We thank  Marco Cirelli, Timothy Cohen, Liam Fitzpatrick,  Ryuichiro Kitano,  Luca Latronico, Yasunori Nomura, Enrico Pajer, Aaron Pierce, Maxim Perelstein, Bibhushan Shakya, Tracy Slatyer, Yuhsin Tsai, and Hai-Bo Yu for discussions and useful comments. 
B.B, C.C. and P.T. are supported in part by the NSF grant PHY-0757868. J.H. and J.S. are supported by the Syracuse University College of Arts and Sciences and by the U.S. Department of Energy under grant DE-FG02-85ER40237. P.T. is also supported by an NSF Graduate Research Fellowship and a Paul \& Daisy Soros Fellowship for New Americans.

\appendix

\section{Explicit models}
\label{app:models}

We present explicit models to demonstrate how one may generate different values of the coupling $b_1$, defined in (\ref{eq:b1:expression}).
In their simplest form, both examples have an unbroken $R$-symmetry which implies that only the $A$-terms generate a mass for the Goldstone fermion. It is straightforward to modify these examples to explicitly break the $R$-symmetry without modifying the structure of these theories. 

\subsection{The simplest example}

We consider a simple variation of the model considered in \cite{Luty:2010vd} with the superpotential $W=yS(\bar{N}N-\mu^2)$. This gives
\begin{equation}
K=f_N^2 e^{(A+A^\dagger)/f}+f_{\bar{N}}^2 e^{-(A+A^\dagger)/f}
\qquad\qquad
 f^2=f_N^2+f_{\bar N}^2\, ,
\end{equation} 
so that the tree-level range for $b_1$ is
\begin{equation}
-1\leq b_1=\frac{f_N^2-f_{\bar{N}}^2}{f_N^2-f_{\bar{N}}^2}\leq1\,.
\end{equation} 

\subsection{An example with $|b_1|\geq 1$}
A perturbative model that may give  $|b_1|\geq 1$ is the following:
\begin{equation}
W= \lambda XYZ-\mu^2 Z+\frac{\tilde{\lambda}}{2} Y^2 N-\tilde{\mu} \bar{N}N \, ,
\end{equation}
where the charges are $q_{Z}=0$, $q_{N}=-q_{\bar{N}}=-2q_{Y}=2q_{X}$ and all couplings and masses are non-zero.
The resulting supersymmetric minimum 
\begin{equation}
f_{X}f_{Y}=\mu^2/\lambda\,,
\qquad\qquad 
f_Z=f_N=0\,,
\qquad\qquad 
f_{\bar{N}}=\tilde{\lambda}\frac{{f_Y}^2}{2\tilde{\mu}}
\end{equation}
gives vanishing $F$-terms
%\begin{align}
%F_{Z}=& \lambda  XY-\mu^2 \\
%F_{Y}=& \lambda XZ+\tilde{\lambda}NY\\
%F_{X}=& \lambda  YZ \\
%F_{N}=&\frac{\tilde{\lambda}}{2} Y^2-\tilde{\mu}\bar{N}\\
%F_{\bar{N}}= &-\tilde{\mu}N\,.
%\end{align}
while the Goldstone chiral multiplet is
\begin{equation}
A=\sum_i \frac{q_i f_i \psi_i}{f}=\frac{q_{Y}}{f}\left(Yf_{Y}-Xf_{X}+2\bar{N}f_{\bar{N}}\right)
\qquad \qquad
f^2=q_{Y}^2\left(f_{Y}^2+f_{X}^2+4f_{\bar{N}}^2\right)\,.
\end{equation}
The corresponding $b_{1}$ at tree-level is given by
\begin{equation}
b_{1}=\frac{1}{f^2}\left(\sum_{i}q_i^3 f_i^2\right)=\frac{-f_X^2+f_{Y}^2+8 f^2_{\bar{N}}}{f_X^2+f_{Y}^2+4 f^2_{\bar{N}}}
\end{equation}
which goes to $b_1\rightarrow 2$ when $f_{\bar{N}}\gg f_{X,Y}$.

\section{Annihilation cross section}
\label{appAnnihilation}

\begin{figure}[h]
  \centering
  \subfloat[$\chi\chi\to gg$]{\label{fig:ann:gg}\includegraphics[]{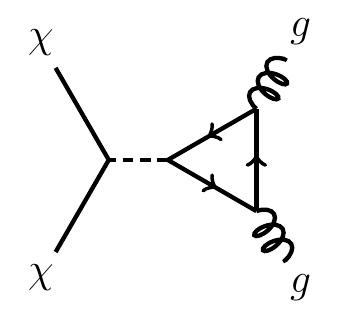}}
	\qquad
  \subfloat[$\chi\chi\to aa$]{\label{fig:ann:aa:t}\includegraphics[]{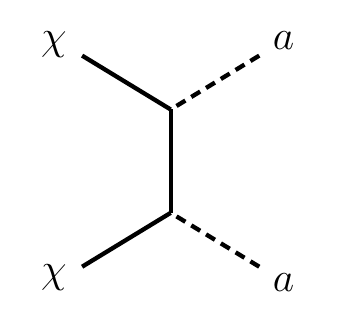}}
	\qquad
  \subfloat[$\chi\chi\to aa$]{\label{fig:ann:aa:u}\includegraphics[]{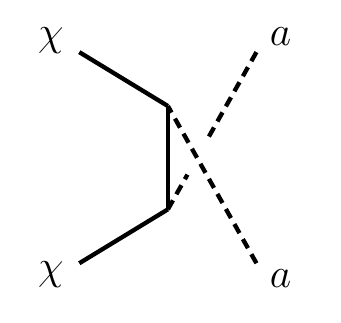}}
\\
\subfloat[$\chi\chi\to aa$]{\label{fig:ann:aa}\includegraphics[]{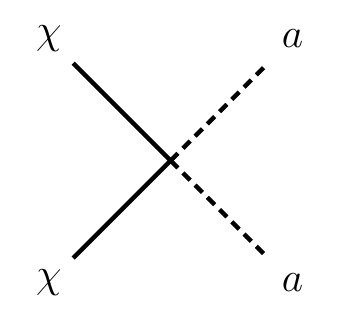}}
	\qquad
\subfloat[$\chi\chi\to ha$]{\label{fig:ann:ha}\includegraphics[]{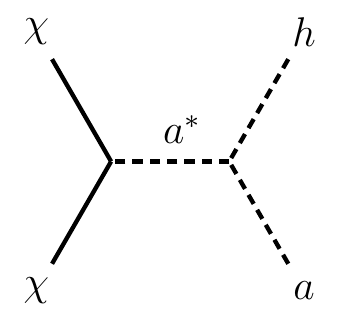}}
	\qquad
\subfloat[$\chi\chi\to hh$]{\label{fig:ann:hh}\includegraphics[]{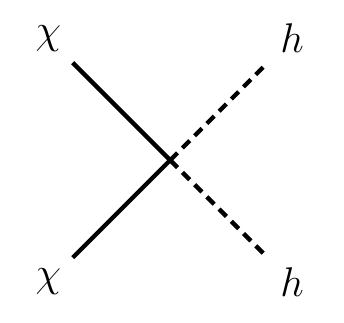}}
  \caption{Goldstone fermion annihilation channels.}
  \label{fig:annihilation}
\end{figure}

Diagrams for the dominant annihilation channels are presented in Fig.~\ref{fig:annihilation}.

\subsection{$\chi \chi \rightarrow gg$}

% \begin{center}
% \begin{tikzpicture}[line width=1.25 pt, scale=.6]
% 	\draw[gluon] (60:1) -- (60:2);
% 	\draw[scalarnoarrow] (180:1) -- (180:2);
% 	\draw[gluon] (300:1) -- (300:2);
% 	\draw[fermion] (300:1) -- (60:1);
% 	\draw[fermion] (60:1) -- (180:1);
% 	\draw[fermion] (180:1) -- (300:1);
% 	\node at (60:2.5) {$g$};
% 	\node at (300:2.5) {$g$};
% 	%
% 	\begin{scope}[shift={(-2,0)}]
% 		\draw (0,0) -- (120:1.8);
% 		\draw (0,0) -- (240:1.8);		
% 		\node at (120:2.3) {$\chi$};
% 		\node at (240:2.3) {$\chi$};
% 	\end{scope}
%  \end{tikzpicture}
% \end{center}

The annihilation cross section to gluons (see Fig.~\ref{fig:ann:gg}) is controlled by the anomalous coupling (\ref{anomaly_int}) where $c_\text{an}=\alpha_s q_{\Psi} N_{\Psi}/(8\pi)$ and the vertex $b_{1}/(2\sqrt{2}f)\bar{\chi }\gamma^{\mu}\gamma^5\chi  \partial_\mu a$.
Away from resonance one finds
\begin{equation}
\sigma v=\frac{2\alpha_s^2}{(8\pi)^3}N_c N_{\Psi}^2 (b_1 m_{\chi }+\delta m_a)^2\frac{s^2 q_{\Psi}^2}{(s-m_a^2)^2f^4}
\qquad \qquad 
s=(p_1+p_2)^2=4 E^2_{\chi }
\label{eq:ggann}
\end{equation}
where $v$ is the relative velocity in the center of mass frame, $N_c=8$ is the number of colors in the final state, and $\delta=-(\alpha+\beta)/2$ is the contribution from the explicit breaking vertex (\ref{new_interactions}).
Note that this process gives a non-vanishing $s-$wave annihilation component.
 
 \subsection{$\chi \chi \rightarrow a a$}

Annihilation into Goldstone bosons proceeds through $t$- and $u$- channel diagrams (see Figs.~\ref{fig:ann:aa:t}--\ref{fig:ann:aa:u}) as well as a contact interaction coming from explicit breaking (see Fig.~\ref{fig:ann:aa}).

 \subsubsection{$t$- and $u$-channel}

These diagrams give a $p$-wave contribution to the cross section, $\sigma v=a+b v^2+\ldots$,
\begin{align}
a=& 0\qquad z=m_a/m_\chi \\ 
b= & \frac{m_\chi^2 }{96\pi f^4(z^2-2)^4} [b_1^3(b_1+4z\delta)(3z^8-16 z^6+48z^4-64z^2+32)\\
\nonumber
&+z^2\delta^2(3z^8-14 z^6+46z^4-64z^2+32)+16b_1\delta^3(z^2-1)(b_1 z^3+\delta(z^2-1))]\, ,
\end{align}
where $\delta=-(\alpha+\beta)/2$ is the contribution from the explicit breaking vertex (\ref{new_interactions}).

\subsubsection{Explicit breaking vertex}
\label{expl_break}

% \begin{center}
% \begin{tikzpicture}[line width=1.25 pt, scale=.7]
% 	\draw[scalarnoarrow] (0,0) -- (45:2);
% 	\draw[scalarnoarrow] (0,0) -- (-45:2);
% 	\draw (0,0) -- (135:2);
% 	\draw (0,0) -- (-135:2);
% 	\node at (45:2.5) {$a$};
% 	\node at (-45:2.5) {$a$};
% 	\node at (135:2.5) {$\chi$};
% 	\node at (-135:2.5) {$\chi$};
%  \end{tikzpicture}
% \end{center}

The quartic contribution to the annihilation cross section is also $p$-wave is
\begin{equation}
\sigma v=\frac{1}{128 \pi}\rho^2 \frac{m_a^2}{f^4} v_a \left(\frac{s-4m_\chi^2}{s}\right)
\qquad 
v_a=\sqrt{1-\frac{4m_a^2}{s}}\, ,
%\rho=\alpha^2+\alpha\beta+\beta^2=7
%\sigma v=\frac{49}{256 \pi}q^4\frac{m_{\chi }^2}{f^4} v^2 v_2=b^\prime v^2 v_2
%\qquad \langle\sigma v\rangle=b^\prime\frac{16}{\sqrt{\pi}x^{3/2}}\,. 
\end{equation}
where $\rho$ is given in terms of the charges of the explicit breaking operators  (\ref{new_interactions}), $\rho=\alpha^2+\alpha\beta+\beta^2$.

\subsubsection{Interference}

The contact interaction interferes with  the $t$- and $u$-channel diagrams.
Summing the amplitudes and then squaring gives,
\begin{equation}
b=\frac{m_\chi^2 b_1^2}{96\pi f^4}(2b_1^2+8b_1 z\delta+z\rho)+\frac{m_\chi^2 z^2}{1536 \pi f^4}\left(3\rho^2+32 b_1\delta\rho+128 b_1^2 \delta^2-16 b_1^4\right)+\mathcal O(z^3)\,, \qquad a=0
\end{equation}
where $\sigma v=a+b v^2+\ldots$ and $z=m_a/m_\chi$. Note that for all plots in this document we use the full expression for $b$ that is valid for all $z\leq 1$.

\subsection{Subleading processes}

The annihilations to a single Higgs (Fig.~\ref{fig:ann:ha}) and to two Higgses (Fig.~\ref{fig:ann:hh}) are subdominant.

\subsubsection{$\chi \chi \rightarrow a^* \rightarrow ah$}

% \begin{center}
% \begin{tikzpicture}[line width=1.25 pt, scale=.8]
% 	% \draw[fermionbar] (-140:1)--(0,0);
% 	% \draw[fermion] (140:1)--(0,0);
% 	\draw (120:1.5)--(0,0);
% 	\draw (240:1.5)--(0,0);
% 	\draw[scalarnoarrow] (0:1.5)--(0,0);
% 	\node at (120:1.9) {$\chi$};
% 	\node at (240:1.9) {$\chi$};
% 	\node at (.9,.4) {$a^*$};	
% \begin{scope}[shift={(1.5,0)}]
% 	\draw[scalarnoarrow] (-60:1.5)--(0,0);
% 	\draw[scalarnoarrow] (60:1.5)--(0,0);
% 	\node at (-60:1.9) {$a$};
% 	\node at (60:1.9) {$h$};	
% \end{scope}
% \end{tikzpicture}
% \end{center}

This channel is available when $2m_{\chi}> m_a+m_h$. Naively, it should be less important because the cross section has an extra suppression by $(v_{\text{EW}}/f)^2$. On the other hand, this is an $s$-wave contribution and therefore the effect is not completely negligible compared to the $\chi\chi\rightarrow aa$ $p$-wave process.
The cross section is given by
\begin{equation}
\sigma v=\frac{v_a}{32\pi}\frac{(b_1 m_\chi+\delta m_a)^2 c_h^2 v_{\text{EW}}^2}{f^6}\left(\frac{m_a^2-m_h^2+s}{s-m_a^2}\right)^2\, .
\end{equation}
% This has a non-vanishing $s$-wave contribution.

\subsubsection{$\chi \chi \rightarrow  hh$}

% \begin{center}
% \begin{tikzpicture}[line width=1.25 pt, scale=.7]
% 	\draw[scalarnoarrow] (0,0) -- (45:2);
% 	\draw[scalarnoarrow] (0,0) -- (-45:2);
% 	\draw (0,0) -- (135:2);
% 	\draw (0,0) -- (-135:2);
% 	\node at (45:2.5) {$h$};
% 	\node at (-45:2.5) {$h$};
% 	\node at (135:2.5) {$\chi$};
% 	\node at (-135:2.5) {$\chi$};
%  \end{tikzpicture}
% \end{center}

This channel is allowed when $m_\chi> m_h$, up to thermal contributions. Because Higgs can be buried under QCD, it is possible to have $m_{h}\sim 90$ GeV.
This process is generated from the contact interaction term $c_{hh}(\bar{\chi}i\gamma^\mu\partial_\mu\chi)h^2/(2f^2)$,
which follows from the $c_2, d_2$ coefficients in the K\"ahler potential.
\begin{equation}
\sigma v=\frac{v_a}{8\pi}\frac{m_\chi^2 c^2_{hh}}{f^4}\left(\frac{s-4m_\chi^2}{s}\right)
\end{equation}
This is a $p$-wave process.

\section{Sommerfeld enhancement from Goldstone boson exchange}
\label{app:Sommerfeld}

\begin{figure}[htb]
\begin{center}
\includegraphics{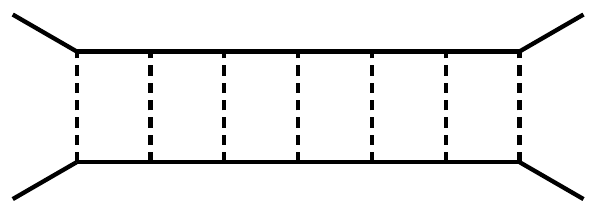}
% \begin{tikzpicture}[line width=1.25 pt, scale=.75]
% 	\draw (0,0) -- (210:1);
% 	\begin{scope}[shift={(0,1.5)}] 
% 		\draw (0,0) -- (150:1);
% 	\end{scope}
% 	%
% 	\draw (0,0) -- (6,0);
% 	\draw (0,1.5) -- (6,1.5);
% 	%
% 	\draw[scalarnoarrow] (0,0)--(0,1.5);
% 	\draw[scalarnoarrow] (1,0)--(1,1.5);
% 	\draw[scalarnoarrow] (2,0)--(2,1.5);
% 	\draw[scalarnoarrow] (3,0)--(3,1.5);
% 	\draw[scalarnoarrow] (4,0)--(4,1.5);
% 	\draw[scalarnoarrow] (5,0)--(5,1.5);
% 	\draw[scalarnoarrow] (6,0)--(6,1.5);
% 	%                               .5
% 	\begin{scope}[shift={(6,0)}] 
% 		\draw (0,0) -- (-30:1);
% 		\begin{scope}[shift={(0,1.5)}] 
% 			\draw (0,0) -- (30:1);
% 		\end{scope}
% 	\end{scope}
% \end{tikzpicture}
 \caption{Exchange of multiple soft Goldstone bosons can lead to an attractive force enhancing the annihilation cross section for the Goldstone fermions.}
\label{fig:Sommerfeld1}
\end{center}
\end{figure}

Thus far we have calculated the relic density assuming no enhancement due to long-range forces. Here we briefly present the non-relativistic potential between the Goldstone fermions and argue that there could be regions of parameter space with a sizeable Sommerfeld enhancement in the annihilation cross section due to an attractive force between the Goldstone fermions due to the exchange of multiple low-energy Goldstone bosons \cite{Bedaque}, as depicted in Fig.~\ref{fig:Sommerfeld1}. It is thereby possible to lower the Goldstone boson and fermion mass scales. 
We emphasize that this enhancement is not necessary to obtain the correct abundance and sufficiently low direct detection cross sections, but it may open up a different region of the parameter space where the Goldstone fermion mass in the $10-50$ GeV range. 

In the non-relativistic limit, the $\chi_1\chi_{2}\to\chi_{1'}\chi_{2'}$ scattering amplitude gives rise to a spin-spin interaction.
The low-energy potential can be written in terms of a traceless tensor and a central piece:
\begin{equation}
V(r)=V_\text{T} (r) \left( 3 \ \vec{S}_1 \cdot \hat{r}\  \vec{S}_2 \cdot \hat{r} - \vec{S}_1 \cdot \vec{S}_2 \right) + V_\text{C}(r) \vec{S}_1 \cdot \vec{S}_2\, ,
\end{equation}
where the coefficients are
\begin{align}
V_\text{T}(r)&= \frac{b_1^2 }{8\pi f^2} \left( \frac{1}{r^3} +\frac{m_a}{r^2} +\frac{1}{3} \frac{m_a^2}{r} \right) e^{-m_a r}  \qquad\qquad
 V_\text{C}(r)&=  \frac{b_1^2 }{8\pi f^2}  \frac{1}{3} \frac{m_a^2}{r} e^{-m_a r}.
\end{align}
Note that the leading term for distances $r< m_a^{-1}$ is contained in the tensor potential. For total spin $S=0$, the tensor potential averages out to zero, whereas the central part gives an attractive interaction  which is independent of the orbital angular momentum
\begin{equation}
\langle S=0\,,\,\ell |V(r)|S=0\,,\,\ell\rangle=-\frac{3}{4} V_\text{C}(r)\,.
\end{equation}
This contribution vanishes in the limit $m_a\rightarrow 0$ in agreement with \cite{ArkaniHamed:2008qn}.
For $S=1$, $\ell=1$ the central potental is repulsive whereas the tensor is attractive. The net effect is an attractive potential\footnote{The potential becomes repulsive for $r m_a \gtrsim 13$. However, this contribution is cut off by the  exponential decay of the Yukawa interaction so that the energy barrier is extremely small $\approx 10^{-11}\times m_a^3/f^2$.}
\begin{equation}
\langle S=1\,,\, \ell=1 |V(r)|S=1\,,\, \ell=1\rangle=\left(\frac{1}{20}-\frac{1}{4}\right) V_\text{T}(r)+\frac{1}{4}V_\text{C}(r)\,.
\end{equation}
The magnitude of this Sommerfeld enhancement was calculated in detail for $s$-wave annihilation processes in \cite{Bedaque}, where it was found to take values as large as 1000 and as small as 0.1.
% It was found that for some regions of parameters it can be as large as 1000, but could also be a suppression. 
For the current model one would only need a factor of few to lower the Goldstone boson and fermion masses to the 10 GeV range. Since most of the leading annihilation channels relevant to this class of models are $p$-wave, the results of \cite{Bedaque} are not directly applicable. A dedicated calculation is left for future work.

\end{document}